\documentclass[sigconf,edbt]{acmart-edbt2023}

\usepackage{libertine}
\usepackage{graphicx}
\usepackage{subfig}  
\usepackage{balance}  
\usepackage{enumitem}
\usepackage{amsmath,amsfonts}

\usepackage{amssymb}
\usepackage{stfloats}
\usepackage{url}
\usepackage[textsize=tiny]{todonotes}
\usepackage{xcolor}
\usepackage{soul}
\usepackage{tabularx}
\usepackage{booktabs}
\usepackage[ruled,vlined,linesnumbered,noresetcount]{algorithm2e}
\usepackage{varwidth}
\usepackage{mathtools}

\def\BibTeX{{\rm B\kern-.05em{\sc i\kern-.025em b}\kern-.08em
    T\kern-.1667em\lower.7ex\hbox{E}\kern-.125emX}}
  
\setcopyright{rightsretained}
\acmDOI{}
\acmISBN{978-3-89318-088-2}
\acmConference[]{EDBT 2023}{}{Ioannina, Greece} 
\acmYear{2022}
\begin{document}

\title{bloomRF: On Performing Range-Queries in Bloom-Filters with Piecewise-Monotone Hash Functions and Prefix Hashing}
\subtitle{[Extended Version]}

\author{Bernhard Mößner, Christian Riegger, Arthur Bernhardt, Ilia Petrov}
\affiliation{
  \institution{Data Management Lab, Reutlingen University}
}
\email{[firstname].[surname]@reutlingen-university.de}

\newcommand{\bloomRFtitle}{\textsf{\textbf{\lowercase{bloom}RF}}}
\newcommand{\bloomRFb}{\textsf{\textbf{\lowercase{bloom}RF}}}
\newcommand{\bloomRFcaption}{\textsf{\textbf{bloomRF}}}
\newcommand{\bloomRF}{{\sf \lowercase{bloom}RF}}
\newcommand{\bloomRFparam}[1]{{\sf \lowercase{bloom}RF(#1)}}
\newcommand{\BFs}{BFs}
\newcommand{\BF}{BF}
\newcommand{\DIs}{DIs}
\newcommand{\DI}{DI}
\newcommand{\code}[1]{\ensuremath{\textit{code}(#1)}}
\newcommand{\codei}[2]{\ensuremath{\textit{code}(#1)_#2}}
\newcommand{\bitshiftr}{\ensuremath{\!>\!>\!}}
\renewcommand{\shorttitle}{\bloomRF{}: Range-Queries in Bloom-Filters with PMHF and Prefix Hashing}
\renewcommand{\shortauthors}{B. Mößner, C. Riegger, A. Bernhardt, I. Petrov}
\newcommand{\tn}{tn}
\newcommand{\tp}{tp}
\newcommand{\fp}{f\!p}
\newcommand{\fpr}{f\!pr}
\newcommand{\pot}{pot}

\begin{abstract}
We introduce \bloomRF{} as a unified method for approximate membership testing that supports both \textit{point-} and \emph{range-queries}. 
As a first core idea, \bloomRF{} introduces novel \textit{prefix hashing} to efficiently encode range information in the hash-code of the key itself. As a second key concept, \bloomRF{} proposes novel  \textit{piecewise-monotone hash-functions} that preserve local order and support fast range-lookups with fewer memory accesses. \bloomRF{} has near-optimal space complexity and constant query complexity.
Although, \bloomRF{} is designed for \textit{integer domains}, it supports \textit{floating-points}, and  can serve as a \textit{multi-attribute} filter. The evaluation in RocksDB and in a standalone library shows that it is more efficient and outperforms existing point-range-filters by up to  4$\times$ across a range of settings and distributions, while keeping the false-positive rate low. 
\end{abstract}

\maketitle

\section{Introduction}
\label{sec:intro}
Modern data sets are large and grow at increasing rates \cite{Gray:SienceExponentialWorld:Nature:2006}. To process them data-intensive systems perform massive scans that incur significant performance and resource consumption penalties. While indices may reduce the scan pressure, they are not always effective due to size or predicate selectivity concerns, or due to the high maintenance costs and workload compatibility.
\emph{Filters} are a class of approximate data structures that may effectively complement the workhorse data structures to reduce scans. \emph{Bloom-Filters (\BFs{})} \cite{Bloom:BF:CACM:1970}, are prominent representatives of this class that are efficient and compact. They avoid false negatives, while false positives are possible, yet the false positive rate (FPR) can be controlled by parameters such as bits/key or the number of hash functions. If a \BF{} returns true, the search key may be present or not and the system needs to verify that through expensive scans or index-lookups. \BFs{} only support point-lookups, i.e. is key 4711 not in the dataset.

{\bf State-of-the-Art Overview.} 
Many algorithms and systems necessitate \emph{efficient range filtering} for queries such as: are there \emph{keys between 42 and 4711} in the dataset. Classical  Prefix \BFs{} or Min/Max indices (fence pointers, ZoneMaps \cite{Neteeza} in Neteeza or Block-Range Index \cite{BRIN} in PostgreSQL) can perform  range-filtering, but are impractical for point queries and result in a higher FPR.

\begin{figure}[!t]
	\begin{center}
		\includegraphics[width=\columnwidth]{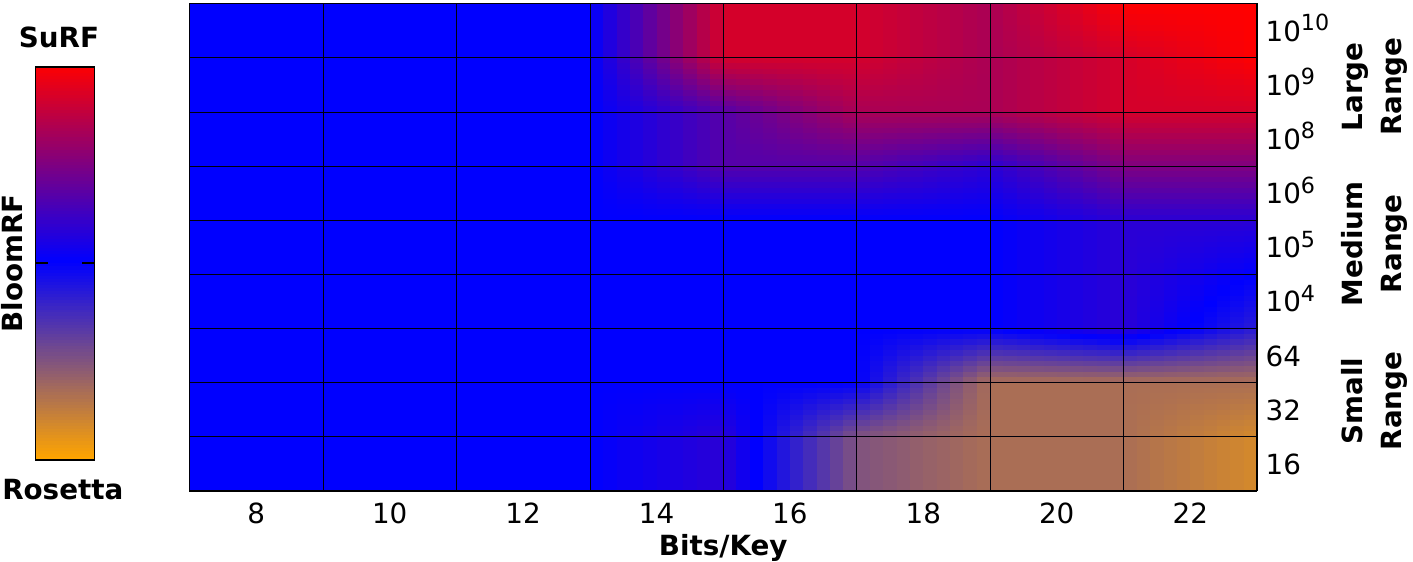}
        \caption{
        \bloomRFb{} is efficient, general and augments existing approaches. The color indicates the approach with the best FPR for different number of keys ($10^{3}$-$5\!\cdot\!10^{7}$), and normal distribution (data/queries) in standalone settings.
        }
		\label{fig:head}
	\end{center}
\end{figure}

Rosetta \cite{Dayan:Rosetta:SIGMOD:2020}, SuRF \cite{Zhang:SURF:SIGMOD:2018} and ARF \cite{Alexiou:ARF:VLDB:2013} are some recent proposals that can handle point- and range-lookups on a unified data structure and serve as \emph{point-range filters} (PRF). ARF \cite{Alexiou:ARF:VLDB:2013} and SuRF \cite{Zhang:SURF:SIGMOD:2018} utilize \emph{tries} and thus partially materialize the index at the cost of extra space. Such techniques result in increased range-filter sizes, that are reduced by trie-truncation or require tedious training/re-optimization. These yield an \emph{a posteriori/offline} creation. 
Rosetta \cite{Dayan:Rosetta:SIGMOD:2020} takes a different approach, where each key is decomposed into a set of prefixes according to a dyadic interval scheme and implicit Segment-Trees \cite{deBerg:SegmentTrees:2008}. The key-prefixes are maintained in a hierarchical set of \BFs{}, one for each prefix length.
%
Fig. \ref{fig:head} shows a holistic PRF positioning in the problem space according to their FPR, for different space budgets and query ranges. It is a flattened version of Fig. \ref{fig:exp3}.E, where we average the FPR for $10^{3}$-$5\!\cdot\!10^{7}$ keys.

\noindent{\bf Problem 1: Existing point-range-filters are designed either for small or for large query ranges}. 
Existing Point-Range-Filters are optimized for handling different query ranges sizes. While Rosetta \cite{Dayan:Rosetta:SIGMOD:2020} excels at relatively small ranges $[2^{1}\!-\!2^{6}]$, SuRF \cite{Zhang:SURF:SIGMOD:2018} offers outstanding FPR for mid- and large-ranges [$2^{37}\!-\!2^{38}]$. On the one hand, as stated in \cite{Dayan:Rosetta:SIGMOD:2020}  trie-truncation techniques, like the ones used in SuRF may lose effectiveness as short query ranges may fall in the scope of the truncated suffixes and thus have higher probability of being detected non-empty. 
On the other hand, range-lookups in Rosetta have logarithmic  (sometimes linear) complexity with respect to the query range size. It may lose efficiency for longer ranges as probing a hierarchical set of \BFs{}, implies higher memory or CPU-costs. Moreover, it is not always possible to bound the query-range size. While short-ranges seem reasonable for KV-stores, this does not apply to other systems or workloads. Besides, datatypes also have an impact: for \emph{doubles} a range of 1 can be $2^{61}$ in the bit representation.

\noindent{\bf Problem 2: Existing point-range-filters are offline.}
Existing PRF (ARF \cite{Alexiou:ARF:VLDB:2013}, SuRF \cite{Zhang:SURF:SIGMOD:2018}, or Rosetta \cite{Dayan:Rosetta:SIGMOD:2020}) employ powerful optimizations, which require \emph{a priori} the complete dataset and are therefore constructed \emph{offline}. Hence, PRF cannot serve range-queries, while data is being simultaneously inserted. 
This limits PRF applicability in the general case, i.e., when PRF are used standalone, when the data is too large, is streamed, or is not available in advance, etc. 

The issue can be mitigated by the way PRF are integrated in larger systems. KV-stores use a main-memory delta area to absorb new data. The PRF leverage that delta and get constructed only when it gets full and thus holds the complete PRF dataset. Searching the main-memory delta is handled otherwise, e.g. through its organization (\texttt{HashSkipLists} or \texttt{HashLinkLists} in RocksDB).
While this approach is practicable in such systems, it: (a) is a property of the system integration, not of the PRF; (b) disregards the extra space for the delta; and (c) is far from optimal in general settings.

Prefix-\BFs{} and Min/Max filters may be constructed online, but are inadequate for point-querying. Rosetta \cite{Dayan:Rosetta:SIGMOD:2020} may be used online per se, yet some of its optimizations require the dataset \emph{a priori}. 

\noindent{\bf Problem 3: Existing Point-Range-Filters exhibit non-robust performance across a variety of workload- and data- distributions.} 
Existing PRF are sensitive to data and workload skew. For instance, Rosetta claims \cite{Dayan:Rosetta:SIGMOD:2020} to outperform SuRF by 2$\times$  on normally distributed workloads in RocksDB \cite{Siying:RocksDB:VLDB20} as the suffix-truncation techniques in SuRF yield more prefix-collisions for small ranges. 

\begin{figure}[b]
	\begin{center}
        \includegraphics[width=\columnwidth]{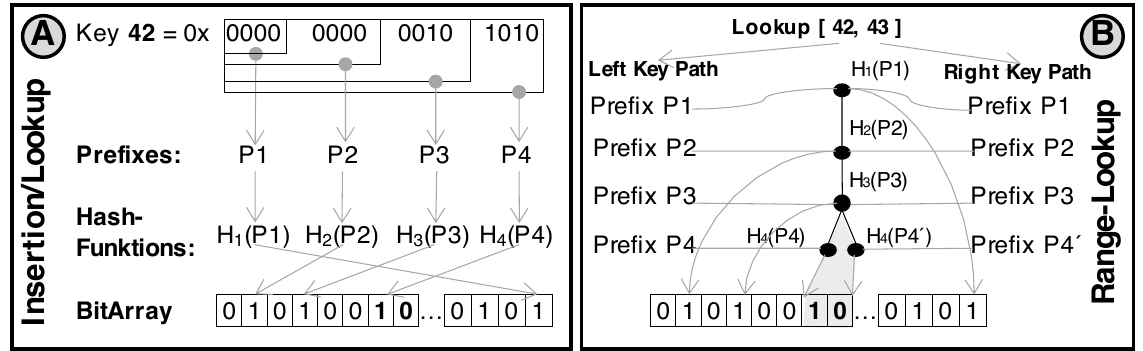} 
        \caption{ (a) \bloomRFb{} relies on PMHF and  prefix hashing. (b) Range lookups traverse two prefix paths, probing automatically the area in between (shaded).}
	\label{fig:headpict}
	\end{center} 
\end{figure}

\noindent{\bf Bloom-Range-Filter (\bloomRFtitle{}).}
We introduce \bloomRF{} as a unified data structure, supporting approximate \emph{point-} and \emph{range-} membership tests that can substitute existing \BFs{}. \bloomRF{} operates on prefixes of keys. Firstly, \bloomRF{} introduces novel \textit{prefix hashing} (Fig. \ref{fig:headpict}) to efficiently encode range information in the hash-code of the key. This information is based on certain dyadic intervals to which the key prefixes correspond. Secondly,  \bloomRF{} proposes novel \emph{piecewise-monotone hash-functions (PMHF)} that preserve local order and support fast range-lookups with fewer memory accesses.
PMHF place information for adjacent prefixes side by side in an overlapped bit-array such that this information can be queried with a single word access. Insertions and point-lookups (Fig. \ref{fig:headpict}.A) behave much like in a \BF{} except that in \bloomRF{} they operate on prefixes. Range-lookups (Fig. \ref{fig:headpict}.B) follow a two-path algorithm, computing the intervals along the prefix-paths for the left and the right key, and probe a tight interval-set. The area in between is probed automatically. PMHF incur fewer memory accesses, e.g. for the query $[42, 43]$ (Fig. \ref{fig:headpict}.B), $H_4$ uses a single  access to probe both points.

Our {\bf contributions} are:
	(a) \bloomRF{} is a unified point-range-filter that is \emph{online} and can serve queries, while data is being simultaneously inserted.
	 (b) \bloomRF{} has \textit{constant query complexity}, independent of the query range size, due to PMHF and its two-path range-lookup algorithm.  \bloomRF{} has a \textit{near-optimal space complexity}, due to prefix hashing.
	(c) \bloomRF{} can serve small-to-large query ranges and can handle different workload- and data-distributions. It supports \emph{integers}, \textit{floating-point numbers}, and can serve as a \emph{multi-attribute} filter.
	(d) \bloomRF{} outperforms all baselines by up to 4$\times$ across a wide range of settings. \bloomRF{} is more \textit{efficient} as it achieves better performance and FPR at lower bits/key.

\noindent{\bf Outline.} 
We continue with a brief background (Sect. \ref{sec:background}), overviewing key terms. On their basis we introduce basic \bloomRF{}'s prefix hashing and PMHF (Sect. \ref{sect:prefix:hashing}, \ref{sect:hash}), and range-lookup algorithm (Sect. \ref{sect:ops}). We present the theoretical model of basic \bloomRF{} and compare it to the theoretical lower bound  \cite{Goswami:RangeEmptiness:SODA15} in Sect. \ref{sect:theoretical:model}, \ref{sect:comparison}. While basic \bloomRF{} is simple, tuning-free, and suitable for ranges $R\!\leq\!2^{14}$, various optimizations (Sect. \ref{sect:optimization}) are needed for larger queries. We present the evaluation in Sect. \ref{sect:eval} and conclude in Sect. \ref{sect:conclusions}.

\section{Background}
\label{sec:background}
We now overview well-known \BFs{} and dyadic intervals from the perspective of \bloomRF{} and establish several key terms.

\noindent\textbf{Bloom-Filters (BF)}.
\label{sec:bloom:filters}
Consider a set
$
  X\!=\!\{x_1,x_2,\ldots,x_n\} \!\subseteq \!D
$
of $n$ elements in a domain $D$ represented by $d$ bits, $|D|\!=\!2^d$, e.g. \emph{d}=16 for UINT16. We call the elements $x \!\in \!X$ keys and arbitrary elements $y\!\in \!D$ lookup keys. A \BF{}\cite{Bloom:BF:CACM:1970} uses a bit-array of \emph{m} bits with positions
$
  M\!=\!\{0,1,\ldots,m-1\}
$
and $k$ hash functions $h_i$ mapping $D$ to $M$ 
(i.e. $
h_i  \colon \! D \! \rightarrow \! M,
i \! = \!k-1,\ldots,1,0
$).
Noticeably, the hash functions transform each lookup key $y \!\in\! D$ in a \underline{\emph{code}} of bit-array positions: 
\begin{equation}
  \label{eq:code:y}
 code(y)
 =
 \big( \; h_{k-1}(y), h_{k-2}(y), \ldots, h_0(y) \; \big) 
 .
\end{equation}
Initially all bits in the bit-array are set to zero. To insert the set of keys $X$ in a \BF{} for each key $x\in X$ the bits of \code{x} are set to one. A \BF{} performs an approximate membership test to decide if a lookup key $y \in D$ is in $X$, by checking, if all the bits of \code{y} are set to one. This procedure may return positive results for elements $y \not \in X$, called \emph{false-positives}. The ratio between false-positives and negatives is called \emph{false-positive rate}.

\noindent\textbf{Dyadic Intervals (DI).}
A \DI{} is an interval whose boundaries are aligned to powers of two. They can be organized in \emph{dyadic levels}, where an interval on level $\ell$ spans $2^\ell$ elements. For a domain represented by $d$ bits there are $d+1$ dyadic levels $\ell \in \{0,1,\dots, d\}$. Each \emph{\DI{}} on level $\ell+1$ is decomposed in two \DIs{} on level $\ell$. Thus \DIs{} form a complete binary tree.
For example, for a domain $D$ of non-negative integers with $d=3$ bits there are $d+1=4$ levels: on level 0 the \DIs{} are the points $[0,0],[1,1]$,\dots, $[7,7]$; on level 1 are $[0,1], [2,3],$ \dots $[6,7]$; on level 2 are $[0,3],[4,7]$; and level 3 has just $[0,7]$. We show how \bloomRF{} encodes \DIs{} with \code{y} in Sect. \ref{sec:bloomRF}.

\noindent\textbf{Prefixes}.
\label{sec:prefixes}
A \emph{prefix} of $y$ on level $\ell$ is the sequence of the $d-\ell$ most significant bits of $y$. These bits are accessed by a right shift by $\ell$ bits (i.e. $y \bitshiftr \ell$), discarding the $\ell$ least significant bits. Thus for $\ell>\ell'$
\begin{equation}
  \label{eq:prfix:hierachy}
  y \bitshiftr \ell = (y \bitshiftr \ell') \bitshiftr (\ell-\ell')
  ,
\end{equation}
i.e., a prefix of $y$ on level $\ell$ is a prefix of a prefix of $y$ on level $\ell'$.

\textit{Noticeably, prefixes are \DIs{}}. There is a one to one correspondence between prefixes on level $\ell$ and \DIs{} on level $\ell$, i.e., all lookup keys $y$ with an identical prefix on level $\ell$ form a \DI{} on level $\ell$.
Consider, for instance, a domain $D$ of non-negative integers represented by $d\!=\!3$ bits. The prefixes of a key $y\!=\!5$ \textit{(bin 0b101)} are $1\!=\!\textit{0b1}$ on level $2$, $2\!=\!\textit{0b10}$ on level $1$ and $5\!=\!\textit{0b101}$ on level $0$. The prefixes of $y\!=\!6\!=\!\textit{0b110}$ are $\textit{0b1}$ on level $2$, $\textit{0b11}$ on level $1$ and $\textit{0b110}$ on level $0$. The prefixes of $y\!=\!7\!=\!\textit{0b111}$ are $\textit{0b1}$ on level $2$, $\textit{0b11}$ on level $1$ and $\textit{0b111}$ on level $0$. Finally, the prefix $\textit{0b11}$ on level $1$ corresponds to the \DI{} $I\!=\![6,7]$ on level $1$. Indeed, exactly the keys $6$ and $7$ share the prefix $\textit{0b11}$ on level $1$.

\section{Bloom-Range-Filter}
\label{sec:bloomRF}
Based on the above concepts we now introduce the main aspects of \bloomRF{} such as \textit{prefix hashing} and \textit{PMHF}.

\begin{table}[b]
	\caption{Most important symbols and abbreviations.}
	\label{tab:abbr}
	\footnotesize 
    \centering
    \begin{tabularx}{\columnwidth}{@{}lXl@{}}
        \toprule
        \midrule
		$D$,$|D|$,$d$   &	domain \emph{D} of size $|D|\!=\!2^{d}$ elements, e.g., $2^{16}$ for UINT16\\		
		$x$, $X$, $n$   &   $x \in X \subseteq D$ - keys in the filter, $|X|=n$ - number of keys\\
		$y$             & 	$y \in D$ - lookup keys\\
		level $\ell$          &   level $\ell \in \{d,\dots, 1,0\}$ - defines prefixes/dyadic intervals of keys\\
		$M$, $m$	    &   $M\!=\!\{0,1,\ldots,m\!-\!1\}$ - bit-array positions, $|M|\!=\!m$ bits\\
		$h_{i}$         &   $h_i \colon D \rightarrow M$ - hash function, $i\in \{k-1,\ldots,1,0\}$\\
		\code{y}        &   a sequence of bit array positions: $\code{y}\!=\! (h_{k-1}(y),\ldots, h_0(y))$\\
        layer $i$             &   layer $i\in \{k-1,\ldots,1,0\}$ - defines prefixes of \code{y}\\
    	\codei{y}{i}    &   prefix of \code{y} on layer $i$: $\codei{y}{i}\!=\! (h_{k-1}(y),\ldots, h_i(y))$\\
		$\ell_i$        &   $\ell_{k-1} \geq \ldots \geq \ell_0$ - level $\ell_i$ corresponds to layer $i$\\
		$\Delta$        &   distance between levels: $\ell_i=i \Delta$\\
		$k$             &   $k\!\approx\!\lceil d/\Delta \rceil$ number of hash functions $h_i$.
		                    Considering $n$ and the saturation of levels
		                    $k\!=\!\lceil (\,d-\log_2\!n\,)/\Delta \rceil$.\\
		$I$,$|I|$       &   $I$ is a lookup interval with $|I|$ elements\\
		$R$             &   upper bound for range-query size: $|I| \leq R$\\
		PMHF & Piecewise-Monotone Hash-Function $M\!H_i$ on layer $i$\\	
		word & bit-array elements of size $2^{\Delta-1}$ bits
		              that PMHF read/write \\
		DI; BF & Dyadic Intervals; Bloom-Filters\\		              
		
        \bottomrule
    \end{tabularx}
\end{table}

\subsection{Prefix Hashing} 
\label{sect:prefix:hashing}
In a \BF{} a lookup key $y$ corresponds to \code{y} of bit-array positions. Thus, we check if $y$ is in $X$ by testing if the bits at \code{y} are set. 

\begin{figure}
\centering
    \includegraphics[width=\columnwidth]{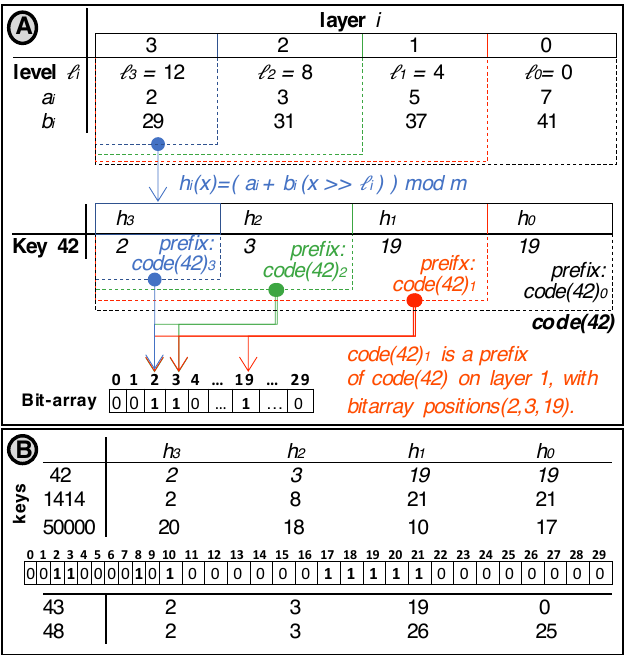}
    \caption{Hash functions and codes of keys.}
    \label{fig:HashfunctionsAndCodes}
    \normalsize
\end{figure}

The core idea of \bloomRF{} is to encode range information in the \code{y} itself. 
To this end, we introduce \underline{\codei{y}{i}} as \textit{the prefix of \code{y} on layer $i$}. We define \codei{y}{i} as an ordered sequence of the first $k\!-\!i$ hash-functions of \code{y}:
\begin{equation}
  \label{eq:prfix:code:y}
  \codei{y}{i}
  =
  \big( \; h_{k-1}(y),\: h_{k-2}(y), \ldots, \: h_{i}(y) \; \big).
\end{equation}
Thus the prefixes \codei{y}{i} are sub-sequences of bit-array positions of \code{y}. As an example we refer to Fig. \ref{fig:HashfunctionsAndCodes}.A, which will be explained in detail below. Here the code of key 42 and prefixes \codei{42}{i} for all layers $i\!\in\!\{3,2,1,0\}$ are shown. 

When performing a lookup, our goal is to check prefixes of lookup key $y$ by testing bits at prefixes of \code{y}. The issue at hand is that there are $d\!+\!1$ dyadic levels, but \code{y} comprises $k$ hash-functions, making it impossible to encode each level. Therefore, we choose to consider only \underline{certain} levels
$
 \ell_{k-1} \geq \ell_{k-2} \geq \ldots \geq \ell_0
 .
$
(Fig. \ref{fig:HashfunctionsAndCodes}.A exemplifies  equidistant levels.)

On this premise, we define \underline{\emph{prefix hashing}} as a key property of \bloomRF{}. It  mandates that for each layer $i\in\{k-1,\ldots,1,0\}$
\textit{a prefix of $y$ on dyadic level $\ell_i$ corresponds to a prefix of \code{y} on layer $i$}, i.e., arbitrary lookup keys $y,y' \!\in\! D$ satisfy
\begin{equation}
  \label{eq:prfix:encoding}
  y \bitshiftr \ell_i = y' \bitshiftr \ell_i \quad \Rightarrow \quad  \codei{y}{i}=\codei{y'}{i}
  .
\end{equation}

Prefix hashing allows using \code{y} to test if \DIs{} on level $\ell_i$ include keys $x \!\in\! X$. Remember that such \DIs{} are prefixes on level $\ell_i$. By prefix hashing such \textit{\DIs{} correspond to prefixes \codei{y}{i}}, which are checked by testing if the bits at \codei{y}{i} are set. This way, \bloomRF{} \textit{implicitly} encodes range information in the \code{y}. 

Arbitrary hash-functions can be used for prefix hashing: eq. (\ref{eq:prfix:encoding}) is satisfied, if hash-functions of \codei{y}{i} only operate on prefixes on level $\geq\! \ell_i$. Using (\ref{eq:prfix:hierachy}) we achieve this by:
\[
 \code{y} \!=\!
 \big( h_{k-1}(y \bitshiftr \ell_{k-1}),
 \ldots,
  h_1(y \bitshiftr \ell_1),
  h_0(y \bitshiftr \ell_0) \big)
\]

Finally, we have to determine the levels $\ell_i$ and the number of hash-functions $k$. A natural choice are equidistant levels. Thus we define a distance $\Delta$ between two adjacent levels and set $\ell_i \!= \! i\Delta$. Then the number of hash-functions is given by $\lceil (d\!+\!1)/\Delta \rceil$. Depending on the number of keys top layers saturate (Sect. \ref{sect:optimization}).
We omit such levels and therefore use $k\!=\!\lceil (\,d-\log_2\!n\,)/\Delta \rceil$ hash-functions.

\noindent\textbf{Introductory example.} Consider a set 
$
 X \!\!=\! \{ 42, 1414, 50000 \} 
$ (Fig. \ref{fig:HashfunctionsAndCodes}.B)
of $n\!=\!3$ keys in a domain $D$ with $d\!=\!16$ bits. We use 10 bits/key and $\Delta\!=\!4$, yielding a bit-array with $m\!=\!10|X| \!=\!30$ bits and $k\!=\!\lceil (d-\log_2\!n)/\Delta \rceil\!=\!4$ hash-functions. For hashing we use multiplication with prime numbers $a_i$ and $b_i$, followed by \texttt{mod} $m$ to determine a position in $M$, i.e.,
$
 h_i(x)\!=\! \big( a_i + b_i \cdot (x \bitshiftr \ell_i) \big) \!\!\!\mod m
 .
$

Figure \ref{fig:HashfunctionsAndCodes}.A shows layers $i$, levels $\ell_i$ and values for the hash-functions. For example, key $42$ has a code $(2,3,19,19)$ of positions in the bit-array. Inserting all keys of $X$ leads to a bit-array where the bits 2, 3, 8, 10, 17, 18, 19, 20 and 21 are set to one and all others are zero (Fig. \ref{fig:HashfunctionsAndCodes}.B).
Since we choose $\Delta\!=\!4$  bit shifts by levels $\ell_i\!=\!4i$ can be displayed in hexadecimal representation. For example, for key 42 (hex \textit{0x002A}) the prefix on level $l_3\!=\!12$ is \textit{0x0}, on level $l_2\!=\!8$ is \textit{0x00}, on level $l_1\!=\!4$ is \textit{0x002} and on level $l_0\!=\!0$ is \textit{0x002A}.

Remember that due to \textit{prefix hashing} (eq. \ref{eq:prfix:encoding}), a prefix of $y$ on  a certain dyadic level $\ell_i$ corresponds to a \codei{y}{i}, i.e. a prefix of \code{y} on layer $i$.
Thus, keys $42$ and $43\!=\!\textit{0x002B}$ have the same prefix on level $\ell_1\!=\!4$, $42 \bitshiftr \ell_1 \!=\! \textit{0x002} \!=\! 43 \bitshiftr \ell_1$ and $\code{42}_1 \!=\! (2,3,19) \!=\! \code{43}_1$ as required by eq.(\ref{eq:prfix:encoding}). 

Recall also that prefix hashing allows us to use \code{y} to test if \DIs{} on level $\ell_i$ include keys $x \!\in\! X$ by testing the positions of \codei{y}{i}, since these \DIs{} are in fact prefixes of $y$ on level $\ell_i$ corresponding to \codei{y}{i}.
All $y \!\in\! I\!=\![32,47]\!=\![\textit{0x0020},\textit{0x002F}]$ have the same prefix \textit{0x002} on level $\ell_1$ and therefore the same prefix $\codei{y}{1}\!=\!(2,3,19)$ on layer 1 (Fig. \ref{fig:HashfunctionsAndCodes}.A,B). Thus we  check positions $(2,3,19)$ to test, if a key $x \!\in\! X$ is included in $I$. In this example, the answer is positive, which is true since indeed $42 \!\in\! I$.
All lookup keys in $[48,63]\!=\![\textit{0x0030},\textit{0x003F}]$ have $(2,3,26)$ as code prefix on layer 1 (e.g. $48$). Checking positions $(2,3,26)$ results negative, since 26 is set to zero, and here indeed $X \cap [48,63] = \emptyset$. 

A \DI{} $I$ on level $\ell$, $\ell_i < \ell < \ell_{i+1}$, can be decomposed in up to $2^{\Delta-1}$ intervals on level $\ell_i$, thus $I$ can be tested via these \DIs{} on level $\ell_i$. For example $I\!=\![42,43]$ on level 1 can be checked by testing $[42,42]$ and $[43,43]$ on level 0. The corresponding prefixes \codei{y}{i} only differ in the hash-function on layer $0$. While $42$ and $43$ are adjacent, the positions of the hash-functions $h_0(42)\!=\!19$ and $h_0(43)\!=\!0$ are not. Clearly, a hash-function on layer $i$ does not preserve the order of the prefixes $y\bitshiftr \ell_i$. We tackle this in Sect. \ref{sect:hash}.

\noindent\textbf{Prefix hashing is hierarchical}.
\DIs{} are arranged hierarchically by inclusion. Prefixes are \DIs{} and follow the same hierarchy -- eq.  (\ref{eq:prfix:hierachy}). The prefixes \codei{y}{i} also inherit that hierarchy by (\ref{eq:prfix:encoding}), hence \bloomRF{} uses hierarchical hashing. Thus, by testing key $y\!\in\! D$, all \DIs{} on levels $\ell_i$ including $y$ are automatically tested. 

For example, when testing key $y\!=\!43$ with $\code{y}\!=\!(2,3,19,0)$, the following prefixes are checked: prefix $(2,3,19)\!=$\codei{43}{1} corresponding to \DI{} $[32,47]$, prefix $(2,3)\!=$\codei{43}{2} corresponding to $[0,255]$ and prefix $(2)\!=$\codei{43}{3} corresponding to $[0,4095]$.

\noindent\textbf{Space Efficiency}. 
\bloomRF{} has a \textit{near-optimal space efficiency} (Sect. \ref{sect:comparison}) since \code{y} itself contains range information in terms of corresponding \DIs{}. In particular, prefix hashing encodes the difference between any two consecutive prefixes of a key \emph{in a single position as a single bit}. For example, the difference between prefixes \textit{0x002} on level 4 and  \textit{0x002A} on level 0 of key 42 is encoded in a single bit.

\subsection{Piecewise-Monotone Hash-Functions}
\label{sect:hash}

Although prefix hashing results in near-optimal space consumption the order of prefixes $y \bitshiftr \ell_i$ is not preserved by hash-function $h_i$, increasing significantly the query time of intervals $I$ on level $\ell$, $\ell_i < \ell < \ell_{i+1}$. To this end, and as a second core idea, \bloomRF{} introduces \textit{piecewise-monotone hash-functions (PMHF)} that are locally order preserving and place corresponding bits side by side in the bit-array. This allows checking all bits of \DIs{} of $I$ on level $\ell_i$ with hash-function $h_i$, in a \textit{single memory access}, yielding better performance.

Noticeably, arbitrary hash-functions $h_i$ can be easily extended to satisfy this property and remain compute-efficient:

$ 
 \Big(
 \Big(
  h_i\big(x \!>\!>\! (\Delta -1) \big) \!\!\! \mod \frac{m}{2^{\Delta-1}}
 \Big) \!<\!<\! (\Delta -1) \Big)
 +  x \,\&\, (2^\Delta-1)
$ 

The new \textit{h} must preserve the order of the least significant $\Delta-1$ bits of a prefix. Therefore, $x$ is right-shifted by $\Delta-1$ bits, such that $h$ only operates on the rest. The bit-array is accessed in \textit{words} of size $2^{\Delta-1}$, therefore $m$ must be a multiple of $2^{\Delta-1}$. In fact, the bit-array can be viewed as an array of $m/2^{\Delta-1}$ words. The modulo operation determines a position in this word-array. Finally a left-shift by $\Delta -1$ bits, yields the position of the word in the bit-array. To keep the order the least significant $\Delta-1$ bits are added to the position. These bits are extracted with a bitwise \textit{AND} ($\&$) with the mask $2^\Delta\!-\!1$.
Combining with prefix hashing we get 
\begin{align*}
 M\!H_i(x) = \,
 &
 \Big(
 \Big(
   h_i\big(x \!>\!>\! (\ell_i+\Delta -1) \big) \!\!\!\!\!\mod \frac{m}{2^{\Delta-1}}
 \Big) \!<\!<\! (\Delta -1)\Big)\\
 &+  (x \!>\!>\! \ell_i) \,\&\, (2^\Delta-1)
 ,
\end{align*}
which we call \textit{piecewise-monotone hash-functions}.

\begin{figure}
\centering
  \includegraphics[width=\columnwidth]{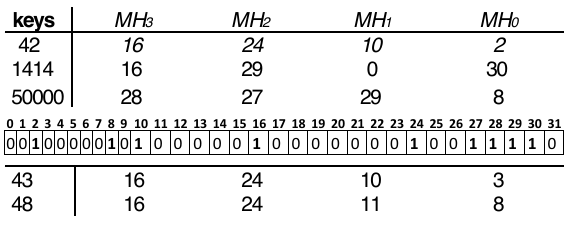}
  \caption{PMHF codes of keys (positions in bit- array)}
\label{fig:PMHFAndCodes}
\normalsize
\end{figure}

For example, consider again the set
$
 X\!=\! \{ 42, 1414, 50000 \}
$ (Fig. \ref{fig:PMHFAndCodes})
for a domain with $d\!=\!16$ bits. Again, we use $\Delta\!=\!4$ and $k\!=\!4$ hash-functions. Here we set $m\!=\!32$ since $m$ must be a multiple of $2^{\Delta-1}\!=\!8$, thus we use approximately 10 bits per key. Again we use the hash-functions
$
 h_i(x)\!=\! a_i + b_ix
$
as in the previous example. Figure \ref{fig:PMHFAndCodes} shows the codes of keys $x \!\in\! X$ using corresponding PMHF. Inserting all keys of $X$ leads to a bit-array where the bits 0, 2, 8, 10, 16, 24, 27, 28, 29 and 30 are set to one and all others are zero.

To test the \DI{} $[42,43]$ the codes $(16,24,10,2)$ and $(16,24,10,3)$ have to be checked. Both have the same prefix 16, 24 and 10 on levels 3 to 1 and the positions 2 and 3 on level 0 lie side by side. Thus on level 0 both can be tested with a \textit{single word access}. The positions 2 and 3 on level $0$ can be described by the bit-mask $b \!=\! \textit{0b \!0011 \!0000}$ and a word access on the first byte of the bit-array yields $w \!=\! \textit{0b \!1010 \!0000}$. The bits at 16, 24 and 10 are set $b \, \& \, w \neq 0$, thus a positive answer.

For interval $[44,47]$ all codes $(16,24,10,4),$\dots, $(16,24,10,7)$ have to be tested. They have the same prefix 16, 24 and 10 on levels 3 to 1 and positions 4 to 7 on level 0 lie side by side and can be tested with a \textit{single word access}. Positions 4 to 7 on level $0$ correspond to the bit-mask $b \!=\! \textit{0b \!0000 \!1111}$ and as above $w \!=\! \textit{0b \!1010 \!0000}$. The bits at 16, 24 and 10 are set, but $b \, \& \, w \!=\! 0$, thus the negative answer.

\begin{figure}[b]
	\centering
  \includegraphics[width=\linewidth]{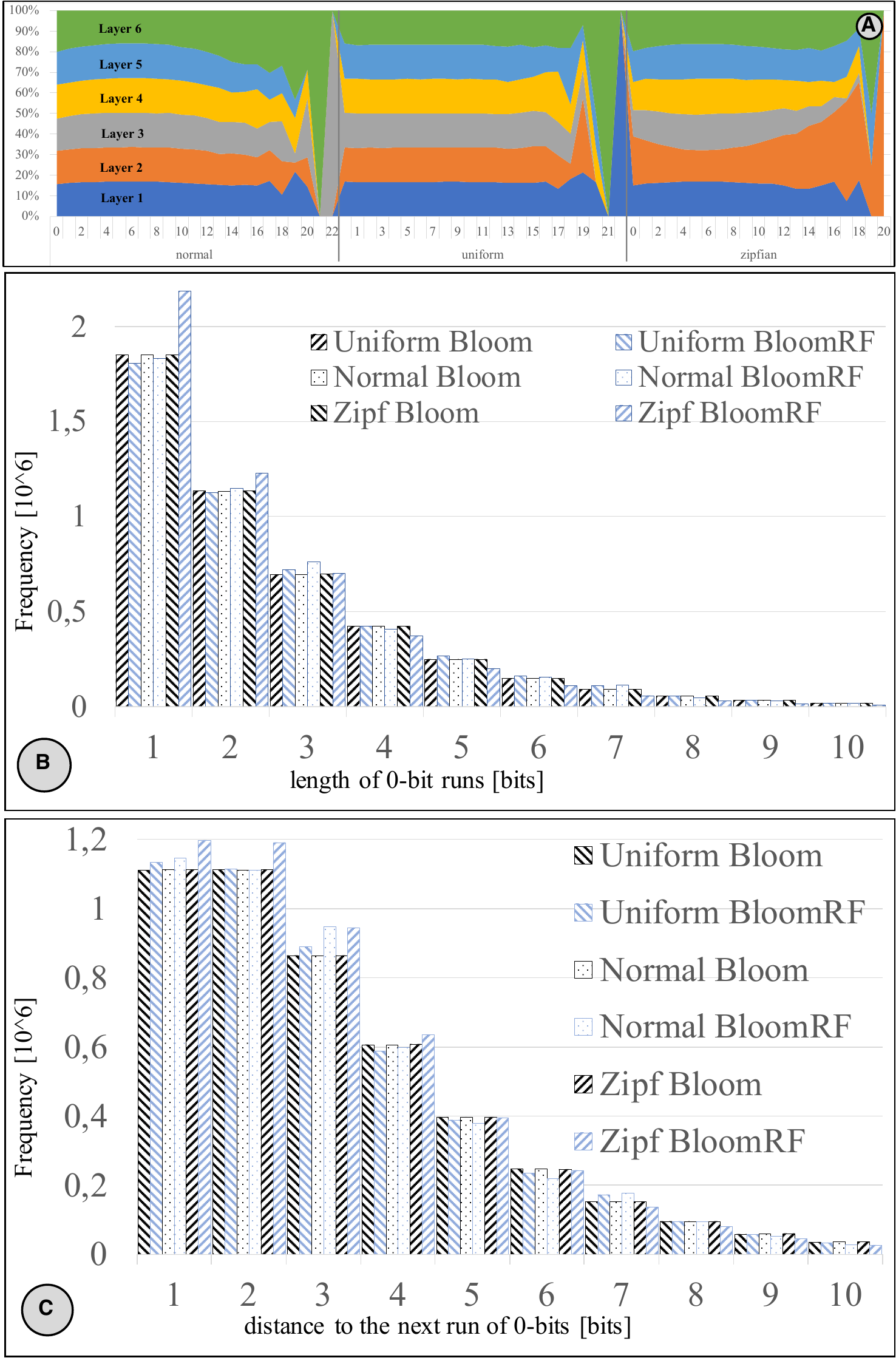}
  \caption{(a) Random scatter over \bloomRFb{} layers; (b), (c) Comparison of the bit-array scatter to a \BF{}.}
  \label{fig:ecc} 	
\end{figure}

\noindent{\bf Random Scatter.} 
We now consider the scatter of PMHF as they should preserve local order, but also distribute words randomly over the bit-array. We compare \bloomRF{} against the standard \BF{} in RocksDB. For a fair comparison we use 2M keys and 10 bits/key, for which \BFs{} have $10\!\cdot\! \ln{2}\!=\!6.93$ hash functions, floored to 6 in RocksDB, as basic \bloomRF{} with 64-bit words ($\Delta\!=\!7$) uses $k\!=\!\lceil (d\!-\!\log_2\!n)/\Delta \rceil\!=\!6$ PMHF. 
\emph{First}, we investigate how well PMHF scatter words. To this end (Fig. \ref{fig:ecc}.A), we measure how many times words (x-axis) of different layers are overlaid in a bit-array element for different data distributions. As the relative frequencies are mostly flat curves (the strong zipfian  skew affects layers 2 and 3) we conclude that PMHF scatter randomly at word granularity for normal, zipfian and uniform data distributions. 
\emph{Second}, we consider the scatter/overlying of bits within words, by looking holistically at the bit-array. To this end, we compare the length of 0-bit runs (Fig. \ref{fig:ecc}.B), as well as the bit-distance between two consecutive 0-bit runs (Fig. \ref{fig:ecc}.C), for both \BF{} and \bloomRF{} and zipfian, normal, uniform data distributions. The 0-bit runs are a relevant metric as they indicate bit areas that have never been set. Thus, significant differences would indicate randomization issues. Clearly, both bit-arrays are in similar states.
\textsf{Intuition:} \bloomRF{} is not worse than \BFs{}, with view of the scatter of  words and their overlaying in bit array elements for common data distributions like zipfian, normal or uniform. PMHF randomize words sufficiently. These insights are substantiated by the relative point FPR of \bloomRF{} vs \BF{} in the evaluation (Fig. \ref{fig:exp2}).

\noindent
\textbf{Degenerate data distributions and PMHF}.
There are rare cases of degenerate data distributions, where PMHF may become inefficient.
The core observation is that  certain bits of a key determine the bit position in a word of the bit-array, since PMHF are piecewise monotone. In basic \bloomRF{} with distance among levels $\Delta=7$, for example, if all bits 0-5, 7-12, 14-19, \ldots, $i\Delta \ldots (i+1)\Delta\!-\!2$, \ldots contain the value $\lambda\in 0,1,\ldots,w=2^{\Delta-1}\!-\!1$, then every PMHF sets bit $\lambda$ in its word. A data distribution that generates such keys with high probability can be defined by counting the number of appearances of these bits in a key 
$
 c_x
 \!=\!
 | \{ i=0,1,\ldots,k-1 \, | \, ((x\!>\!>\! i\Delta) \, \& \, w) = \lambda\}|+1
$
and finally normalizing 
$
 p_x = c_x/\sum_y c_y
$.
\bloomRF{} can  handle such cases. We can employ slightly different hash functions, which permute the bits in the word. 
For instance, on each layer, we can apply the original PMHF on \textit{half} of the keys, while the other half is tackled by a PMHF that writes the words in \textit{reverse order}.

\begin{figure}[t]
	\begin{center}
		\includegraphics[width=\columnwidth]{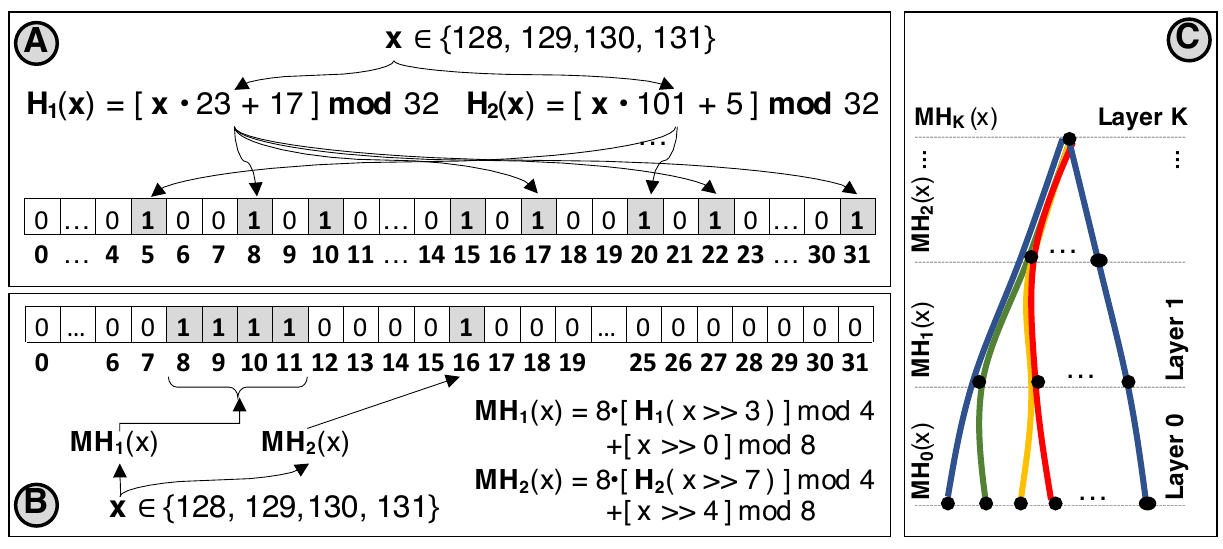}
		\caption{
		(a) Hashing in \BFs{} is planar.  
		(b) PMHF  preserve partial order assisting range-querying, and are
		(c) hierarchical.
		}
		\label{fig:hashfunc}
	\end{center}
\end{figure}

\noindent\textbf{Vertical PMHF and error-correction}.
The hierarchical structure of PMHF allows a new interpretation of hashing in \bloomRF{}, which can also be transferred to \BFs{}. \DIs{} on a level $\ell_i$ correspond to prefixes \codei{y}{i}. With PMHF only the hash-function on layer $i$ operates on $y \bitshiftr \ell_i$, all others only on prefixes of $y \bitshiftr \ell_i$. Thus one hash-function is primarily responsible for level $\ell_i$, namely $M\!H_i$. Therefore hashing in \bloomRF{} is \emph{hierarchical} with a separate PMHF for each layer (Fig. \ref{fig:hashfunc}.C). The hash-functions on higher layers are used for \emph{error-correction}.
In the example (Fig. \ref{fig:PMHFAndCodes}) the \DI{} $I\!=\![416,431]$ an level $\ell_1\!=\!4$ is represented by the prefix $(16,25,2)$. We have $I\cap X \!=\! \emptyset$, but hash-function $M\!H_1$, which is primarily responsible for layer $1$, yields an error, since the bit at position 2 is set to 1. Hash-function $M\!H_2$ checks bit 25 of the bit-array, which is zero. Thus, we get a negative as the error of $M\!H_1$ is corrected by $M\!H_2$.
\BFs{} can be viewed in the same way: Keys are represented by one hash-function while the others are used for error-correction. Since hashing in \BFs{} is planar and not hierarchical (Fig. \ref{fig:hashfunc}.A,B), none of the hash-functions is preferred for representing keys or error-correction.  

\section{\bloomRFb{} Operations}
\label{sect:ops}

We now provide a detailed description of the main operations in \bloomRF{} such as insertion, point- or range-queries.

\noindent{\bf Insertion and Point-Lookup.}
To insert a key $x\!\in\! X$ the \code{x} of bit-array positions is computed via piecewise-monotone hash-functions $M\!H_{i}(x)$, $i\!=\!k\!-\!1,\ldots,1,0$, and the corresponding bits in the bit-array are set to one. To test, if a lookup key $y\in D$ is in $X$, the \code{y} of bitarray positions is computed via PMHF $M\!H_{i}(y)$, $i\!=\!k\!-\!1,\ldots,1,0$, and \bloomRF{} checks if all corresponding bits in the bit-array are set.
For these operations \bloomRF{} behaves like a regular \BF{}, except that the hash-functions are replaced by PMHF.

\noindent{\bf Range-Lookup.}
Range-queries in \bloomRF{} are based on the decomposition of arbitrary lookup intervals $I$ in \DIs{}. Hierarchical prefix hashing allows testing all these intervals together in one pass. Additional \DIs{} covering $I$ are automatically checked. Next, we explain, which \DIs{} are considered for an interval $I$. Upon that we elaborate on the algorithm that computes and tests all these \DIs{}.

\emph{Decomposition in \DIs{}.} For an arbitrary interval $I$ the \DIs{} to be considered are defined in a two-path algorithm, one for the left and one for the right bound of $I$. Starting from the top level, $J_d\!=\!D$ is a covering of $I$. We proceed recursively. Suppose $J_{\ell+1}$, $\ell\!<\!d$, is a covering of $I$. Then we decompose $J_{\ell+1}\!=\!K^l \!\cup\! K^r$ in two \DIs{} and set $J_\ell$ as the one covering $I$. If $I$ is not covered by a single \DI{} the path of covering intervals splits in two, a left $J_\ell^l\!=\!K^l$ and a right $J_\ell^r\!=\!K^r$.

\begin{figure}[b]
    \centering
	\includegraphics[width=\columnwidth]{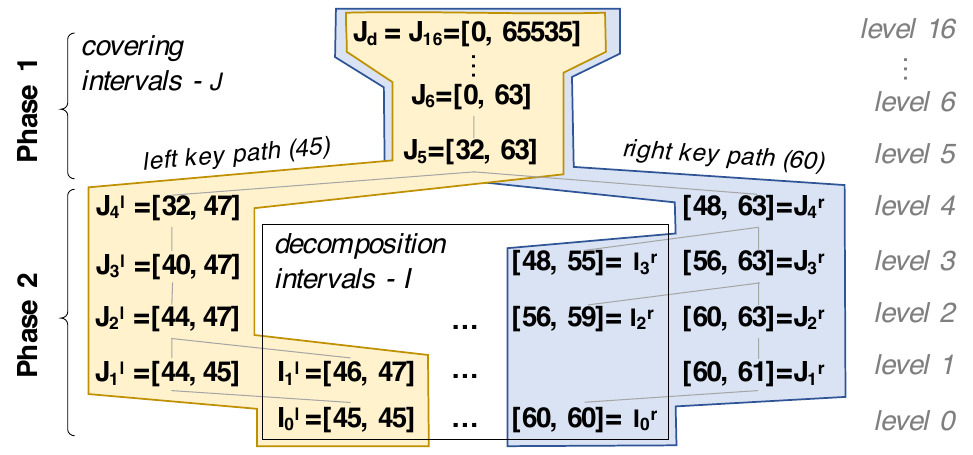}
    \caption{Dyadic intervals for range-query of I=[45,60].}
    \label{ExampleDyadicIntervalsRangeQuery}
\end{figure}

We describe only \DIs{} considered for the left path as the right one is mirror-inverted. Suppose $J_{\ell+1}^l$, $\ell\!<\!d$, is a covering of the left bound of $I$. We decompose $J_{\ell+1}^l\!=\!K^l \!\cup\! K^r$ in two \DIs{}. If $K^l \cap I \!\neq\! \emptyset$, we know that $K^r \!\subseteq\! I$, thus $I_\ell^l\!=\!K^r$ belongs to the decomposition of $I$ in \DIs{} and $J_\ell^l\!=\!K^l$ is a covering of the left bound of $I$. Else, if $K^r \cap I \!\neq\! K^r$, then $K^r$ covers the left bound of $I$, thus we set $J_\ell^l\!=\!K^r$. Otherwise the decomposition of the left side is complete and we set $I_\ell^l\!=\!K^r$.
As example we look at the considered \DIs{} for $I\!=\![45,60]$, $d\!=\!16$  (Fig.  \ref{ExampleDyadicIntervalsRangeQuery}). From level 16 to 5 $I$ is covered by single \DIs{}. On the top levels  left and right path coincide. On level 4 the paths split with a covering of $I$ by two \DIs{}. On level 3 the first \DI{} $I_3^r$ of the decomposition of $I$ is calculated. Finally, $I\!=\![45,45] \cup [46,47] \cup [48,55] \cup [56,59] \cup [60,60]$.

Next, we map the above intervals onto the layers. A covering $J_{\ell_{i+1}}^l$ on level $\ell_{i+1}$ is split into several \DIs{} on the levels $\ell_i \! \leq \! \ell \!<\! \ell_{i}\!+\!\Delta$. These can be represented by at most $2^\Delta$ \DIs{} on level $\ell_i$. Some of them are coverings, while others belong to the decomposition of $I$. All \DIs{} of the decomposition have to be tested. The covering should be as tight as possible, thus we take the intersection of the intervals $J_\ell^l$, which is $J_{\ell_i}^l$. Using PMHF $2^{\Delta-1}$ \DIs{} an level $\ell_i$ lay side by side in the bit-array, thus all \DIs{} to be tested can be checked with at most two word-accesses. The same applies to the right path, thus checks require at most four word-accesses per layer.
In our example the decomposition of $I\!=\![45,60]$ results in intervals $[45,47]\!=\!I_0^l \cup I_1^l$, $[48,55]\!=\!I_3^r$ and $[56,60]\!=\!I_2^r \cup I_0^r$ to be probed. 
Thereby, the coverings are automatically checked:  $J_{\ell_4}\!=\!J_{16}\!=\![0,65535]$ on level 16, $J_{\ell_3}\!=\!J_{12}\!=\![0,4095]$ on level 12, $J_{\ell_2}\!=\!J_{8}\!=\![0,255]$ on level 8 and on level 4 a covering with $J_{\ell_1}^l\!=\!J_{4}^l\!=\![32,47]$ and  $J_{\ell_1}^r\!=\!J_{4}^r\!=\![48,63]$.

\begin{algorithm}[t]
\caption{\bloomRFb{} Range-Lookup}
\label{alg:rangeq}
\small
	\SetKwFunction{FRangeQ}{RangeLookup}
	\SetKwProg{Pn}{Function}{:}{\KwRet}
	\Pn{\FRangeQ{ \textit{l\_key, r\_key} \  } }{
		Let $i \gets$ $k-1$\;
	    \label{alg:rq:line:init} 
		Let $\textit{checks} \gets$ \textbf{init\_checks}(l\_key, \ r\_key, \ i)\;
		\label{alg:rq:line:start:check}
		\While{$\textit{checks} \neq \emptyset$} {
		\label{alg:rq:line:layerloop}
		    Let $\textit{new\_checks} \gets \emptyset$\;
		    \ForEach{$\textit{check} \!\in\! \textit{checks}$}{
		    \label{alg:rq:line:checksloop}
    			\If{$\textit{check.is\_covering} = \textit{true}$} { 
		    		\If{$\textit{filter.bit\_access}(MH_i(\textit{check.l\_key})) = 1$} {
			    	\label{alg:rq:line:check:covering} 
				    	Expand \textit{check} to layer $i-1$ an append to \textit{new\_checks}\;
				    }
			    }
    			\Else{
	    		
		    	    Let $b\! \gets\! \textbf{bit\_mask}(check.l\_key, \ check.r\_key)$\;	
			        \label{alg:rq:line:decomposition}
			        Let $w\! \gets\! \textit{filter.word\_access}(MH_i(\textit{check.l\_key}))$\;
    			    \If{$b \, \& \, w \neq 0$} {
		    	\KwRet{$true$}\;
			    	}
		    	}
		    }
			Let $i \gets$ $i-1$\;
		    Let $\textit{checks} \gets \textit{new\_checks}$\;
		}
		\KwRet{\textit{false}}\;
	}
\end{algorithm}

\emph{Detailed algorithm.} We now describe how \bloomRF{} performs range queries for arbitrary intervals $I\!=\!\textit{[l\_key,r\_key]}$ (Algorithm \ref{alg:rangeq}). The main loop  iterates over the layers, with $i$ being the current layer, which ranges from the top $i\!=\!k\!-\!1$ (Line \ref{alg:rq:line:init}) down to the bottom $i\!=\!0$. On layer $i$ several tests are performed using PMHF $M\!H_i$. The algorithm checks coverings ($J_{\ell_i}$) and intervals of the decomposition of $I$ (unions of intervals $I_{\ell_i}$). The variable \textit{checks} (L. \ref{alg:rq:line:start:check}) contains the data for these tests: \textit{check.l\_key}, \textit{check.r\_key} and \textit{check.is\_covering}. The algorithm loops over the \textit{checks} of layers $i$ (L. \ref{alg:rq:line:checksloop}).
For a covering (L. \ref{alg:rq:line:check:covering}), only a single bit must be tested. If this bit is set, the checks for the underlying layer are computed. Otherwise this interval does not contain any keys $x \in X$. As an early stop condition no further layers have to be checked.  
To test an interval of the decomposition of $I$ (L. \ref{alg:rq:line:decomposition}), the algorithm has to test several bits. We compute a bit-mask $b$ and since we use PMHF, all necessary bits are read in a single-word bit-array access. If a bitwise \textit{AND} yields a value $\neq 0$, then the filter claims the existence of a key $x\in I$ and returns a positive answer. 
Otherwise, if all intervals of the decomposition get excluded, \textit{checks} gets empty, yielding a negative answer.

\section{Theoretical Model}
 \label{sect:theoretical:model}
We now analyze space and time complexity of \bloomRF{}, and begin with an FPR estimate for range-queries. As shown in Sect. \ref{sect:ops}, for an interval \textit{I=[l\_key,r\_key]} several \DIs{} are considered. There are several special cases, depending on the position of $I$. All have in common that they (phase1, Fig. \ref{ExampleDyadicIntervalsRangeQuery}) start with a sequence of $i_1$ coverings by single \DIs{} $J_{\ell_j}$, $j\!=\!k,k\!-\!1,\ldots,k\!-\!i_1\!+\!1$, which then (phase 2) split up in $i_2$ coverings by two \DIs{} $J_{\ell_j}^l \cup J_{\ell_j}^r$, $j\!=\!k\!-\!i_1,k\!-\!i_1\!-\!1,\ldots,k\!-\!i_1\!-\!i_2\!+\!1$.
Since here all intervals are coverings  only single bits have to be checked. Let $p$ be the probability that a bit in the bit-array is set to zero.
A false positive can only occur, if all \DIs{} of phase 1 yield positive and all \DIs{} of left side of phase 2 yield positive, while on the right side an arbitrary combination is possible, or vice versa.
We estimate the FPR $\epsilon$ by eq. (\ref{eq:fpr:est}).
\begin{eqnarray}
  \epsilon
  &\leq&
  (1-p)^{i_1+i_2}(
   2 \sum_{i=0}^{i_2-1}
       \left( \begin{array}{c} i_2\\i \end{array} \right)
       p^{i_2-i}(1-p)^i
     \, + \,
     (1-p)^{i_2}
  )\nonumber\\
  & = &
  2 (1-p)^{i_1+i_2}(
     \underbrace{
     \sum_{i=0}^{i_2}
       \left( \begin{array}{c} i_2\\i \end{array} \right)
       p^{i_2-i}(1-p)^i
     }_{=1}
     \, - \,
     (1-p)^{i_2}
  )\nonumber\\
  & \leq &
  2 (1-p)^{i_1+i_2} \label{eq:fpr:est}
\end{eqnarray} 

The \DIs{} on level $\ell_i=i\Delta$ have length $2^{i\Delta}$. Thus, an arbitrary interval $I$ of length $|I|\leq 2^{i\Delta}$ is covered by at most two \DIs{} on level $\ell_i$ and therefore in phase 2 at least layer $i$ is reached, i.e. $i_1+i_2 \geq k-i$.
Thus $\epsilon \leq 2 (1-p)^{k-i}$ and therefore $\epsilon \!\leq\! 2 (1-p)^{k-log_2(|I|)/\Delta}$.

It remains to estimate the probability $p$ that a bit in the bit-array is set to zero. For \BFs{} the assumption of perfect random hash-functions leads to a probability of $1/m$ of bits being set and therefore the standard estimate \cite{Bloom:BF:CACM:1970} yields
\[
  p = \left(1-\frac{1}{m}\right)^{kn} \approx e^{-\frac{kn}{m}}
  .
\]
We model the influence of the data distribution on PMHF by introducing a constant $C$, such that $p=(1-C/m)^{kn} \approx e^{-Ckn/m}$. Our experiments (PMHF random scatter, Fig. \ref{fig:ecc}) suggest that $C=1$ for common distributions such as uniform, normal and zipfian. 
\textsf{Summary:} for range lookups with max. query range $R$, such that $|I| \!\leq\! R \!=\! 2^{i\Delta}$, and common distributions \bloomRF{} has an FPR of
\begin{equation}
  \label{FPRbloomRF}
  \epsilon
  \leq
  2 \left( 1-e^{-\frac{kn}{m}}\right)^{k-\log_2(R)/\Delta},
\end{equation}
where $k\!=\!\lceil (d-\log_2 \! n)/\Delta \rceil$ (Sect. \ref{sect:prefix:hashing}).

For point-queries \bloomRF{} behaves like a  \BF{}, except that $k$, the number of hash-functions, is not a free parameter. Thus for common distributions the point FPR is $\epsilon \approx (1-e^{-\frac{kn}{m}})^k$.
 
For \textit{time complexity} we consider the operations in Sect. \ref{sect:ops}. The insertion of keys and point lookups requires evaluating of $k$ hash-functions, thus both have constant time $\mathcal O (k)$.
Range-queries are handled by algorithm \ref{alg:rangeq}. There are two loops: The outer loop (Line \ref{alg:rq:line:layerloop}) iterates over the layers and the inner loop (Line \ref{alg:rq:line:checksloop}) over the checks on layer $i$. Since there are $k$ layers and at most 4 word-accesses per layer, range-queries require also constant time $\mathcal O (k)$. Notably, the query-time is independent of the size of the query-interval $I$.

\section{Comparison: space/time complexity}
\label{sect:comparison}
With the theoretical model in place, we now compare \bloomRF{}'s space and time complexity to \emph{Rosetta}'s model, and to the theoretical lower bounds for point\cite{Carter:ExactApproxMembershipTests:STOC:1978} and range-queries\cite{Goswami:RangeEmptiness:SODA15}. 

\noindent{\bf Space complexity.}
We estimate the space $m$ needed by \bloomRF{} to achieve a given FPR $\epsilon$ by solving eq. (\ref{FPRbloomRF}) for $m$.

\begin{figure}[t]
	\centering
  \includegraphics[width=\columnwidth]{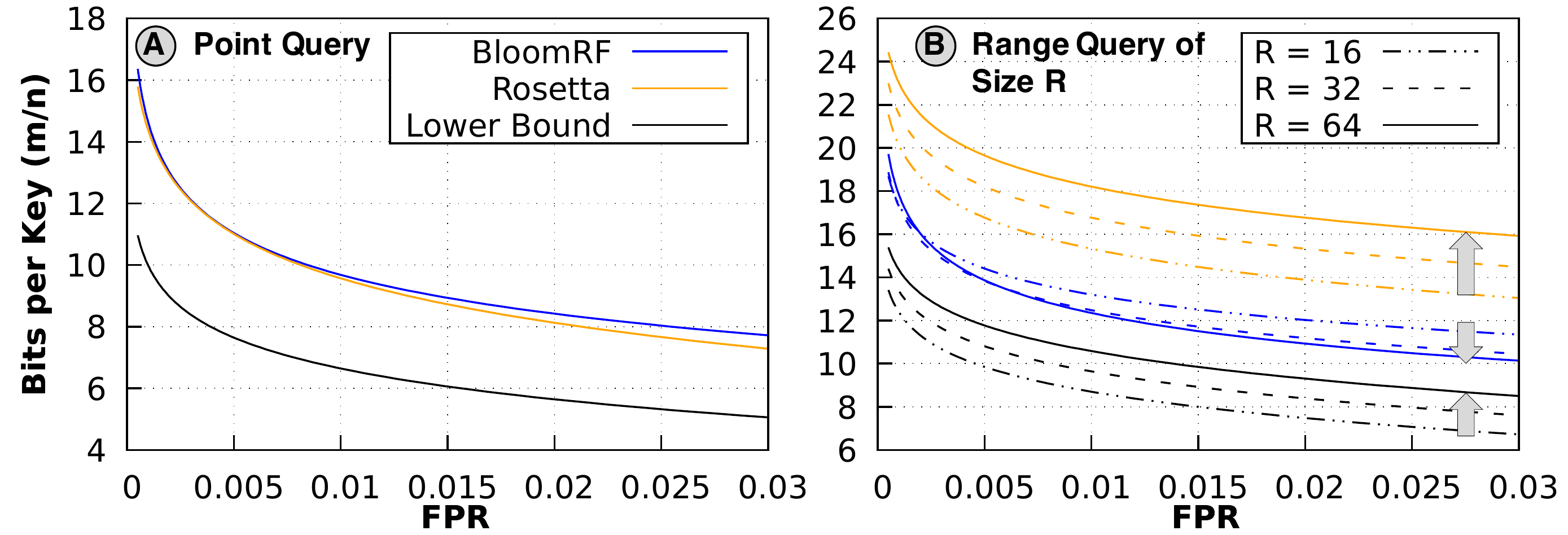}
  \caption{Comparison of \bloomRFb{} to Rosetta and the theoretical lower-bound \cite{Carter:ExactApproxMembershipTests:STOC:1978,Goswami:RangeEmptiness:SODA15} for (a) point and (b) range lookups.}
  \label{fig:fpr:comparison} 	
\end{figure}

\cite{Carter:ExactApproxMembershipTests:STOC:1978} has shown that any structure which answers point-queries with FPR $\epsilon$, needs at least $m \geq n \log_2(1/\epsilon)$ space.  \cite{Goswami:RangeEmptiness:SODA15} shows that any structure, answering range-queries of range-size $R$ with FPR $\epsilon$, necessitates at least $m \geq n \log_2(R^{1-O(\epsilon)}/\epsilon)-O(n)$ space.
\cite{Goswami:RangeEmptiness:SODA15} gives a family of lower bounds with a free parameter, $\gamma>1$:
\[
  m
  \geq
  n \log_2 \left( \frac{R^{1-\gamma\epsilon}}{\epsilon} \right)
  +
  n \log_2 \frac{
    \left(1-\frac{4nR}{2^d}\right)
    \left(1-\frac{1}{\gamma}\right)
  }{e}
  .
\]
The lower bound is therefore the point-wise maximum of these bounds. We can determine $\gamma$ as a function of $\epsilon$ to achieve this maximum, leading to a single curve for the lower bound (Fig. \ref{fig:fpr:comparison}).  

Furthermore we compare with \emph{Rosetta}~\cite{Dayan:Rosetta:SIGMOD:2020}, which has four variants for point-range filters and the variant (F) first-cut solution in analyzed in terms of space complexity. The first-cut solution uses a \BF{} for each level of \DIs{}, with FPR of $\epsilon$ on the bottom level and $1/(2-\epsilon)$ on all others.
In \cite{Dayan:Rosetta:SIGMOD:2020} it is stated, that (F) achieves an FPR of $\epsilon$ for range-queries of intervals up to length $R$ using $m \approx \log_2(e) \cdot n \log_2(R/\epsilon)$.
Figure~\ref{fig:fpr:comparison} shows the estimates for \bloomRF{}, lower bounds and \emph{Rosetta} for point-queries (left) and range-queries of intervals of length $R=16, 32, 64$ (right) for a domain of $d=64$ bit integers.
For point-queries \bloomRF{} and \emph{Rosetta} are close, but \bloomRF{} always uses a little bit more memory except at one FPR. The reason is, that for \bloomRF{} the number of hash-functions is determined by the datatype's domain size, $k\!=\!\lceil (d-\log_2\!n)/\Delta \rceil$, such that the for \BFs{} known optimal choice of $k=\ln(2) \cdot m/n$ cannot be used. 
For range-queries the distance between \emph{Rosetta} an the lower bound is given by a near-constant factor. \bloomRF{} improves over \emph{Rosetta}, especially with larger $R$, i.e., larger $\Delta$, and gets closer to the theoretical lower bound.
The foundation for space savings in \bloomRF{} is \emph{prefix hashing}, yielding a near space-optimal PRF.

\noindent{\bf Time-Complexity.}
Range queries in \bloomRF{} are answered in constant time $\mathcal O (k)$, independent of the range size $R$.

\emph{Rosetta} uses a \BF{} for every dyadic level, but all levels except to lowest have larger FPRs, e.g.  $1/(2-\epsilon)$ in the first-cut solution (F). To improve FPR a process of doubting is applied. If a \DI{} on level $\ell$ yields a positive result, the two \DIs{} on the level below are tested. In the worst case, this may yield query-time linear in $R$. According to  \cite{Dayan:Rosetta:SIGMOD:2020}, (F) has avg. query time $\mathcal O (\log_2(R)/\theta^2)$ for intervals $\leq\!R$.

Two more variants also have log. avg. query-time $\mathcal O (\log_2(R)/\theta'^2)$ \cite{Dayan:Rosetta:SIGMOD:2020}. An optimized variant (O), where as in (F) a \BF{} is used for each level, but the FPRs $\epsilon_\ell$ on the levels are adjusted to how often intervals are queried, and a variable-level variant (V) similar to (O), but using different weights, pushing more bits to lower levels, improving FPR of lower at cost of higher FPR of the middle and top levels.
Finally a single-level variant (S) is suggested, where only a single level of \emph{Rosetta} is used. Here range-queries are answered by testing every element of an interval, yielding linear time \cite{Dayan:Rosetta:SIGMOD:2020}.

\noindent{\bf Space efficiency, FPR and Query-range size.}
Rosetta is designed for relatively small query ranges and KV-stores, and covers its design space well. Consider Rosetta (F) due to its logarithmic time complexity for longer ranges and its space requirement of $log_{2}(e)\cdot nlog_{2}(R/\epsilon)$ bits/key to achieve an FPR $\epsilon$ for range-queries $R\approx 2^\ell$ \cite{Goswami:RangeEmptiness:SODA15,Dayan:Rosetta:SIGMOD:2020}. 
For example, to achieve an FPR  of $2 \%$ for ranges $|R|\!=\!2^6$, Rosetta uses 17 bits/key, yet for $|R|\!=\!2^{10}$ it already demands 22 bits/key, while for $|R|\!=\!2^{14}$ it requires 28 bits/key. 
Given 17 bits/key, basic \bloomRF{} can handle ranges of $|R|\!=\!2^{14}$ with an FPR of 1.5\% and low probe-latency, while with 22 bits/key basic \bloomRF{} covers $|R|\!=\!2^{21}$ with 2.5\% FPR, whereas with  optimizations \bloomRF{} can improve on those significantly.

\section{Optimizations}
 \label{sect:optimization}
\textbf{Observation.} Basic \bloomRF{} is simple, tuning-free, and can handle range queries with $R\!\leq\!2^{14}$ with acceptable FPR and space budgets. However, the  theoretical model also shows that further optimizations for larger $R$ are needed. Next, we describe them. 

\noindent{\bf Replicated Hash-Functions.}
The number of hash-functions to query \DIs{} $I$ decreases with $|I|$. Larger \DIs{} correspond to shorter prefixes, by prefix-hashing to shorter prefixes of the \textit{code} and thus less hash-functions are used, weakening error-correction.
To increase the number of hash-functions on higher layers \bloomRF{} uses \emph{replicated hash-functions}. They write replica of words of the original PMHF $M\!H_i$ but at different bitarray word-positions, preserving the local order defined by PMHF $M\!H_{i}$. Thus \bloomRF{} has $r_i$ functions per layer (incl. $M\!H_{i}$), where typically $r_i\!=\!1$ for lower layers.

\noindent{\bf Variable Distance Between Levels.}
While large $\Delta$ between levels work well on lower layers, as basic \bloomRF{} shows, for higher layers the exponentially increasing size of \DIs{} is one reason for the rapid increase of the FPR. Therefore we aim at smaller distances on higher layers. To this end \bloomRF{} uses a vector $\Delta\!=\!(\Delta_{k-1},\Delta_{k-2},\ldots,\Delta_0) \in \mathbb N^k$, defining level $\ell_i$ corresponding to layer $i$ by $l_i\!=\!\sum_{j=0}^{i-1} \Delta_j$. Smaller distances on higher layers also increase the number of hash-functions on higher layers.

\noindent{\bf Memory Management.}
The relative frequency of \DIs{} including keys changes with the level. On low levels this frequency is nearly zero.  Consider for example $n=50 \cdot 10^6$ keys in a domain of integers with $d=64$.  Level $0$ is nearly empty, since $n<\!\!<2^d$. But each increment of the level halves the number of intervals. Thus in mid levels more intervals are occupied and top levels saturate, depending on the data distribution. Hash-functions used on saturated levels almost always yield positive answers, such that these functions can be omitted. In the above example, levels 39 to 64 (26 top levels) saturate, given a uniform key distribution.

The next lower levels are more then 25\% occupied, but their size is not much larger then $n$. For example level 38 has size $\approx\!1.34 n$, level 37 has size $\approx\!2.68 n$. Therefore a radical design decision is to use one segment to store one level of \DIs{} in an exact bitmap. 

From eq. (\ref{FPRbloomRF}) we see the FPR decreases step-wise with the level. For better balance, we  adjust the probability $p$, by separating the bit-array into $S\!=\!3$ memory segments, one for an exact layer, one for the mid layers and one for the lower layers. More formally $m\!=\!m_1\!+\!m_2\!+\!\ldots+m_S$, where layer $i$ is assigned a segment $j_i \in \{1,2,\ldots, S\}$. The size $m_1$ is determined by the position of the exact layer. By increasing $m_2$, the segment for the mid layers, we can improve $p$ for these layers, thus improving the FPR an larger intervals, simultaneously reducing $m_3$ and thus FPR on the lower layers, especially for point-queries.

\noindent{\it Summary:}
To handle large query ranges \bloomRF{} typically employs the following strategy: 
(i) sparser bottom layers with large word sizes (e.g. 64-bit) are packed together in one segment of the bit-array with a single PMHF per layer;
(ii) mid-layers with small word sizes (e.g. 8-bit or smaller) are stored in a separate and sparser segment with replicating hash-functions besides the PMHFs to lower the error-rates;
(iii) a mid-upper layer is stored exactly in an exclusive segment;  
(iv) the top layers are discarded as they saturate.

\noindent{\bf Extended Model.}
\label{sect:param}
We now describe a general \bloomRF{} model to evaluate the FPR given the above optimizations. According to the filter the \DIs{} on each level $\ell$ can be classified as: (a) empty ($\tn_\ell$); or are (b) non-empty and include a key ($\tp_\ell$); or are (c) non-empty and do not include a key ($\fp_\ell$).
Therefore, the FPR on level $\ell$ is $\fpr_\ell = \fp_\ell/(\fp_\ell+\tn_\ell)$.

The number of true positives $\tp_\ell$ on each level can be derived from the distribution of the keys. For example, assuming uniform distribution, the $n$ keys lie in approximately $n$ \DIs{} on large enough levels. Hence, the estimate: $\tp_\ell=\min(n,2^{d-\ell})$.
The numbers $\fp_\ell$ and $\tn_\ell$ are estimated by recursion on the levels $l_i\!=\!\sum_{j=0}^{i-1} \Delta_j$ corresponding to layers $i$. We assume level $\ell_k$ is stored exactly, therefore
$\fp_\ell\!=\!0$ and $\tn_\ell\!=\!2^{d-\ell}\!-\!\tp_\ell$, $\ell=d,d\!-\!1,\ldots, \ell_k$.
 
Suppose we have computed $\fp_{\ell_i}$ and $\tn_{\ell_i}$ corresponding to layer $i$. For the layer below, i.e., layer $i-1$, we consider the levels $\ell=\ell_i-1, \ell_i-2, \ldots, \ell_{i-1}$. A \DI{} on a level splits in two \DIs{} on the underlying level.   Therefore, each \DI{} $I$ on level $\ell_i$ includes $2^{\ell_i-\ell}$ \DIs{} on level $\ell$.
If $I$ is true negative, all $2^{\ell_i-\ell}$ intervals are also true negatives.
If $I$ is false or true positive, then some of the $2^{\ell_i-\ell}$ intervals can be false positive. Since $\tp_\ell$ are true positive the number of potentially false positive intervals on level $\ell$ is
$
 \fp^{\pot}_\ell
 \!=\!
 2^{\ell_i-\ell}(\fp_{\ell_i}+\tp_{\ell_i})\!-\!\tp_\ell
 .
$

For these intervals the corresponding bits in segment $j_{i-1}$ of the bit-array will be probed. Let $p$ be the probability that such a bit is set to zero. Analogous to section \ref{sect:theoretical:model} we use the estimate
$
 p
 \!=\!
 ( 1\!-\! C/m_{j_{i-1}})^{k' \cdot n}
 ,
$
where $C$ models the influence of the data distribution and here $k'\!=\! \sum_{j_\nu=j_{i-1}} \!r_\nu$ is the number of hash functions of segment $j_{i-1}$. For common distributions such as uniform, normal and zipfian we can assume $C\!=\!1$ (PMHF random scatter, Fig. \ref{fig:ecc}). 

For each potentially false positive \DI{} on level $\ell$ one or more bits will eventually be probed, depending on the number of hash-functions $r_{i-1}$ and layer $i-1$. Let $p'$ be the probability that such a probe yields true, then $\fp_\ell\! = \! p' \fp^{\pot}_\ell$ and $\tn_\ell\! =\! 2^{\ell_i-\ell} \tn_{\ell_i} + (1-p')\fp^{\pot}_\ell$.
The probability  $p'$ can be computed by combinatorial formulas.
For example for \DIs{} on level $\ell_{i-1}$ single bits a checked for each hash-function. Hence, $p'=(1-p)^{r_{i-1}}$. For \DIs{} on level $\ell_{i-1}\!+\!1$ two bits must be checked. For $r_{i-1}\!=\!1$ we get $p'=2p(1-p)+(1-p)^2$, for $r_{i-1}\!=\!2$ we get
$
 p'
 =
 2p^2(1-p)^2
 +
 4p(1-p)^3
 +
 (1-p)^4
 ,
$ etc.

We apply the FPR-model to our example in Section \ref{sec:bloomRF}. The size of the domain is $|D|\!=\!2^{d}\!=\!16384$, $d\!=\!16$, and we store $n\!=\!3$ keys. We assume $\Delta\!=\!4$ and thus $k\!=\!\lceil (\,d-\log_2\!n\,)/\Delta \rceil=4$, or $\Delta\!=\!(4,4,4,4)$. We also assume one hash function per layer and a single shared segment, which is the bit-array with $m_1\!=\!32$ bits. Level $\ell_4\!=\!d$ is the interval $[0,16384]$, which is set when the first key is inserted. Thus, we assume it is stored exactly (it is a single bit, which is actually unused). 
In our model we estimate $p \approx 0.683$, where the relative frequency of bits set to 0 is $22/32 \approx 0.688$. As estimate for the FPR on each level we get $\fpr = (0,0.95,0.78,0.53,0.32,0.27,\ldots,0.04,0.03,0.02,0.01)$. So for point-queries we expect an FPR of 0.01 ($1\%$) and for the intervals $[0,32767], [32768,65535]$ an FPR of $0.95$ ($95\%$).

\noindent{\bf Tuning Advisor.}
Given standard parameters like the number of keys $n$, the memory budget $m$ and considering an (approx. max.) query range size $R$, the tuning advisor computes and selects an appropriate \bloomRF{} configuration, comprising the parameters: vector $\Delta\!=\!(\Delta_{k-1},\Delta_{k-2},\ldots,\Delta_0) \!\in\! \mathbb N^k$, number of hash-functions $r_i$ and the assigned memory segment $j_i$ per layer, while using three segments $(m_1,m_2,m_3)$. Now we describe the procedure. 

First, we determine the exact level by means of a heuristic: its size should be $\leq60\%$ of the memory budget $m$.
Thus, $\ell_e \!=\! \min \{ \ell \, | \, 2^{d-\ell}\!< \!0.6m \}$.
The advisor examines multiple exact level candidates.
For the sake of simplicity, here we consider only:
$\ell_e$ and $\ell_e\!+\!1$.

The position of the exact layer determines the vector $\Delta$, the number of hash-functions and the assigned memory segments by the following heuristics: For the lower layers we use $\Delta_i=7$, which leads to a word-size of 64 bit and is as large as possible. The mid layers are the transition region between lower layers and exact level. Starting from the lower layers we reduce $\Delta_i$ to match the exact layer.
As an example we consider $n=50 \cdot 10^6$ keys with 14 bits/key in a domain with $d=64$ bit. The lowest level with $2^{d-\ell}\!<\!0.6m$ is 36. For the bottom levels we start with $\Delta_i=7$ and then reduce $\Delta_i$ to match 36. This results in a vector $\Delta=(2,2,4,7,7,7,7)$, which sums up to 36. We aim for as few  replicated hash-functions as possible, therefore we use only one hash-function per layer, and only on the highest layer 2, e.g., $r\!=\!(2,1,1,1,1,1,1)$. The heuristic applied here is: the closer we are to the exact layer, the higher the precision has to be, and  therefore we employ smaller $\Delta_i$ and use replicated hash-functions (but as few as possible).
Finally memory segment $m_1\!=\!2^{d-l}$ is used for the exact, $m_2$ for the middle and $m_3$ for the bottom layers, e.g., $j\!=\!(2,2,2,3,3,3,3)$. 

\begin{figure}[h]
	\centering
  \includegraphics[width=\columnwidth]{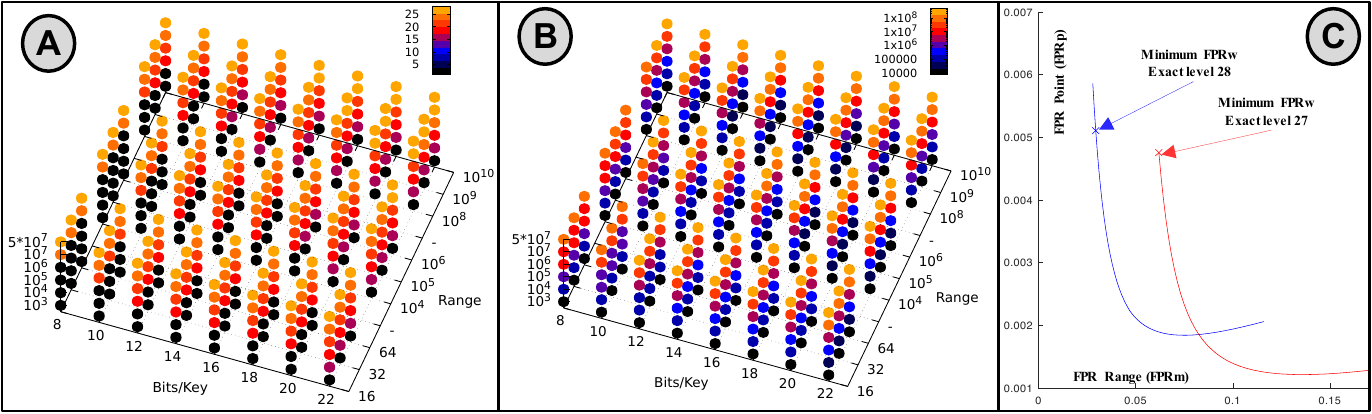}
  \caption{ Sensitivity analysis of the tuning advisor: (a) exact layer $\ell_0$; (b) size of $m_2$.  (c) Estimates for $\fpr_m$ vs. $\fpr_p$.}
  \label{fig:EO} 	
\end{figure}

Second, with all other parameters defined by the above heuristics, for a given exact level it remains to determine $m_2$, since $m_1\!+\!m_2\!+\!m_3\!=\!m$. The goal is to minimize the FPR for range-queries of size up to $R$. Let $\fpr_m \!=\! \max_{\ell\!=0}^{\lfloor \log_2(R) \rfloor} \fpr_\ell$ be the maximum FPR of \DIs{} used for ranges $\leq R$. Since the largest FPR-rates result from mid-top levels (= large intervals), small intervals (= bottom levels) are under-prioritized. Thus we also consider  $\fpr_p\!=\!\fpr_0$, i.e. point-query FPR. The advisor makes a trade-off between lowering the   range-query FPR ($\fpr_m$) and the point-query FPR ($\fpr_p$), as decreasing $\fpr_m$ might imply higher $\fpr_p$. To this end, we define and minimize the weighted squared norm $\fpr_w^2=\fpr_m^2 + C^{2} \fpr_p^2$. It always holds $\fpr_p \!\leq\! \fpr_m$. As compensation we can increase $C$ to weight point-queries stronger.
We determine all parameters for our exact level candidates $\ell_e$ and $\ell_e-1$ and select the configuration with min. $\fpr_w$. 
Finally, we select the configuration with minimum $\fpr_w$.
The auto-tuning process is inexpensive, with computation times of  \textasciitilde 8ms.  
Figure \ref{fig:EO}.C shows an example.
For $n\!=\!50$M keys, 16 bits/key and query range $|R|\!=\!10^{10}$,
the advisor examines $\ell_e\!=\!27$ (red curve) and $\ell_e\!=\!28$ (blue curve).
The minimum $\fpr_w$ is marked on
each curve and the blue one is chosen.
Thus, we estimate an FPR of
\textasciitilde $0.5 \%$ for point-queries and
\textasciitilde $3 \%$ for dyadic ranges up to size $|R|$.

\section{Datatype Support}
\label{sect:datatypes}
 
\noindent{\bf Variable-length strings.} 
The string support in \bloomRF{} resembles SuRF-Hash \cite{Zhang:SURF:SIGMOD:2018} and considers the first seven characters in the seven most-significant bytes. In addition, for point queries it computes a one-byte hash-code of the rest of the string, including the length, and places it in the least significant byte. This way \bloomRF{} achieves a UINT64 representation of variable length-strings.

\noindent{\bf Floating-Point Numbers.} 
Floating-point numbers are represented with $q$ bits for the mantissa $\mu$,
$r$ bits for the exponent $e$ and one bit for the sign $s$.
For a bit combination $x$ the represented value is
$fl(x) = s \cdot \mu \cdot 2^e$.
The bit combinations $x$ are ordered as binary numbers.
Since floats are signed, this order is reversed
for negative numbers and is therefore lost.
To this end, we use a map $\varphi$ with
$\varphi(x) = x + 2^{q+r}$ if $x_{q+r}=0$ and
$\varphi(x) = \overline{x}$ (bitwise inverse) otherwise,
which is a monotone coding, i.e.,
$\varphi(x) < \varphi(y) \Leftrightarrow fl(x) < fl(y)$.
For all operations, we use $\varphi(x)$ instead of $x$.
To insert $x$ into \bloomRF{}, we insert $\varphi(x)$.
For a point-query of $x$ we test $\varphi(x)$.
For a range-query $[x,y]$, we perform a range-query
with  $[\varphi(x),\varphi(y)]$.

\noindent{\bf Multi-Attribute \bloomRFtitle{}.} 
The ability to filter on multiple attributes simultaneously is necessary for complex operations in interactive analytics, scientific packages, IoT and AI. \bloomRF{} supports two-dimensional filtering with reduced precision.  To this end we \emph{concatenate} the attribute-values and insert them in \emph{both} combinations. For instance, \bloomRF\textit{(A,B)} will concatenate the values of \textit{A} and \textit{B}, and insert them as tuples \textit{$<$A,B$>$} and \textit{$<$B,A$>$}. The increased space-requirements are lowered by reducing the precision of \textit{A} and \textit{B}, e.g.  to a 32-bit integer. As a result \bloomRF{} can answer queries such as \textit{A$<$42 \!AND\! B=4711}, \textit{A=42 \!AND\! B$>$4711} or \textit{A=42 \!AND\! B=4711}.

\begin{figure*}[!ht]
	\begin{center}
 		\includegraphics[width=\textwidth]{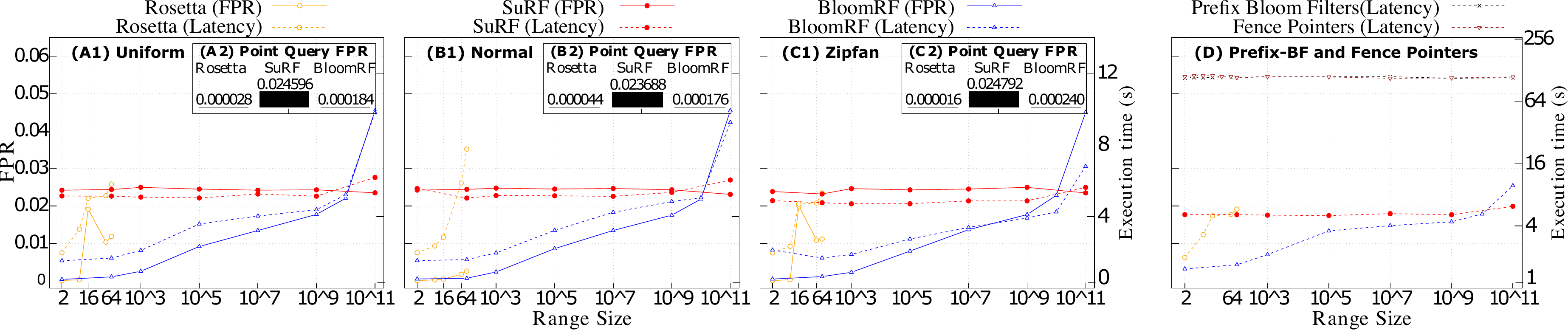}
	    \caption{\bloomRFb{} has good performance for a variety of ranges and workload distributions in RocksDB (22 bits/key).}
		\label{fig:exp1}
	\end{center}
\end{figure*} 

\section{Experimental Evaluation}
\label{sect:eval}
\noindent{\bf Integration in RocksDB \cite{Siying:RocksDB:VLDB20}.} 
\bloomRF{} has been implemented in a standalone library and has been integrated in RocksDB v6.3.6 through a standard filter policy. The policy is extended to pass query-range information (lower/upper bounds) to the filter by means of \textit{slice structures}. For persistence we implement our own ser./deserialization mechanism, placing it as regular \textit{full filter block} in each compaction-disabled SST file of a \textit{block-based table format}.

\noindent{\bf Baselines.} 
Throughout the evaluation the following baselines are used: \BFs{}, Prefix-\BFs{} and fence pointers as well as state-of-the-art point-range filters such as SuRF \cite{Zhang:SURF:SIGMOD:2018,surf:lib,surf:rocksdb} and Rosetta \cite{Dayan:Rosetta:SIGMOD:2020}. We perform two types of experiments. First, \emph{system-level} experiments, where all baselines are compared in RocksDB v6.3.6 to stress the overall effects in a real system. Second, \emph{standalone} experiments are performed to stress specific aspects in isolation.

\noindent{\bf Workloads.} 
Throughout the evaluation we use a set of different workloads. Firstly, we employ a derivative of YCSB \cite{ycsb:2010} Workload E, which is range-scan intensive. The dataset comprises 50M 64-bit integer keys, while the values are 512 bytes long. The data is uniformly distributed, while the workloads are of normal, uniform and zipfian distributions. We issue $10^{5}$ queries of a single  \emph{fixed} range-size that is specified in the respective experiments. All point- and range-queries in this workload are \emph{empty} (unless specified otherwise), which represents the \textit{worst-case}. Depending on the workload, non-empty queries may perform better, e.g. due to \bloomRF{}’s early stop conditions. In fact, in a perfect system a perfect filter would incur minimal I/O, and thus the \textit{worst-case} may overstate their impact. 

Rosetta and \bloomRF{} rely on parameter tuning methods that compute the proper filter-configurations, for given space budgets, number of keys and range sizes. SuRF, however, requires a suffix-length parameter setting to tune itself to a space budget and trade off FPR, by selecting the appropriate variant. 
For some settings, we were unable to select one, especially in RocksDB.
Secondly, for the floating point experiments we use a timeseries dataset from NASA\cite{Kepler:2016}. 
Whereas for the multi-attribute experiments we utilize a dataset from the Sloan Digital Sky Survey DR16 \cite{sloan:SDSS:2019}.

\noindent{\bf Experimental Setup.} 
The experimental server is equipped with an Intel E5-1620 3.50GHz, 32GB DDR4, and runs Ubuntu 16.04.

\noindent{\bf Experiment 1: \bloomRFcaption{} is general-purpose and can handle various query ranges, from large to small.} 
We begin by comparing \bloomRF{} against SuRF and Rosetta in RocksDB under conditions \textit{favorable} to all approaches.
To this end, we employ a space budget of 22 bits/key, 50M uniformly distributed keys and vary the query range sizes and workload distributions (Fig. \ref{fig:exp1}.A1, B1 and C1). 

\begin{figure*}[!ht]
	\begin{center}
\includegraphics[width=\textwidth]{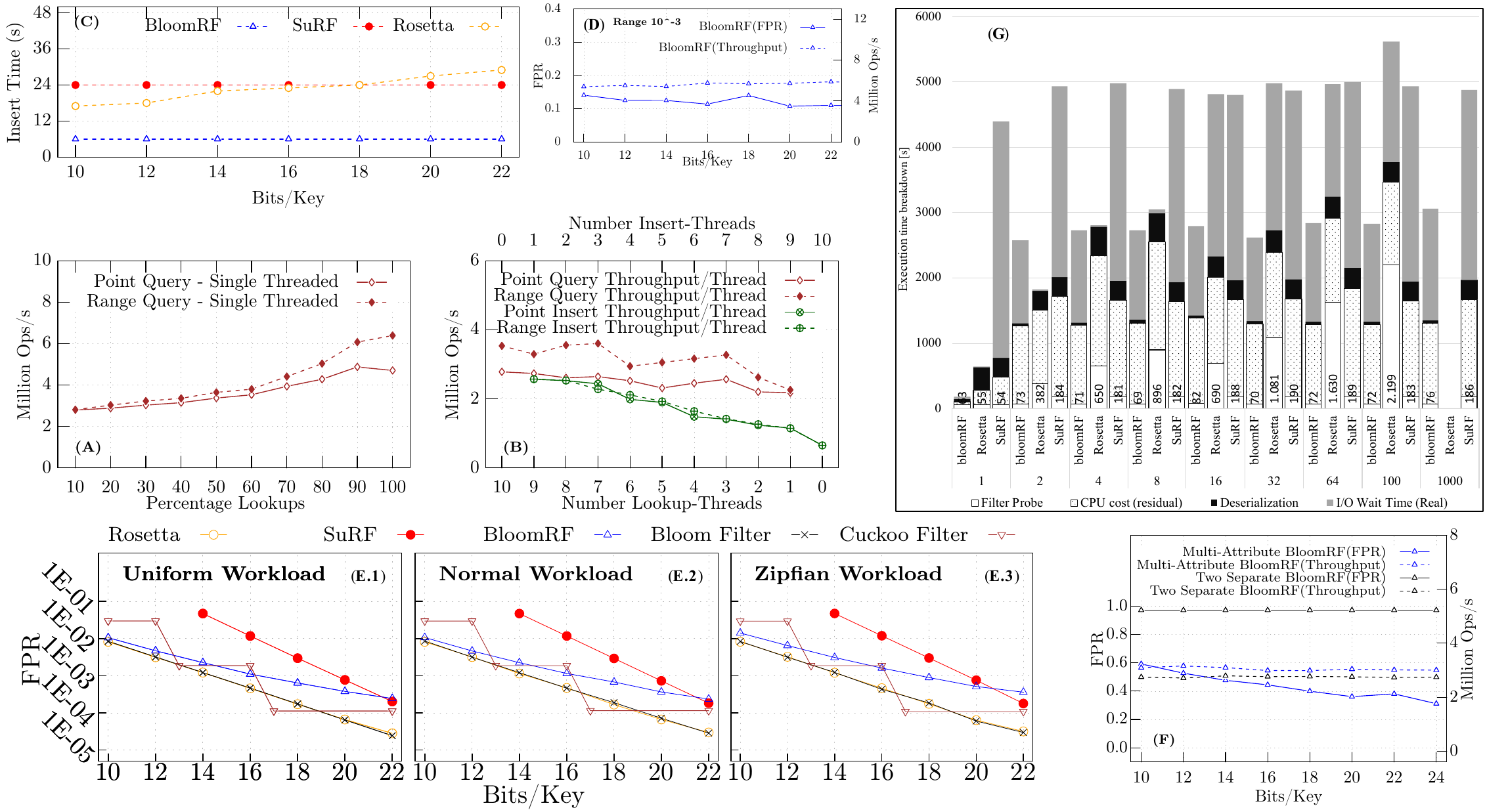}
    \caption{(a, b) online behavior; (c) filter creation; (d) floats; (e) point-queries; (f) dual-attribute filter; (g) cost breakdown.}
    \label{fig:exp4:7}
    \end{center}
\end{figure*}

In terms of \emph{end-to-end probe latency},  \bloomRF{} outperforms all baselines, due to its two-path range-lookup and its CPU-efficient PMHF (Fig. \ref{fig:exp4:7}.G). The sudden rise in \bloomRF{} latency at $|R|\!=\!10^{11}$ is due to approx. 1\% non-empty ranges generated by the workload driver because of the large interval size.
Overall, \bloomRF{} also has the lowest \emph{FPR} of all baselines.
Rosetta is more accurate for \emph{very short} ranges ($|R|\!\leq$ 8) as they hit its precise lower \BF{}. Due to the error-correcting effect of its PMHF \bloomRF{} is more accurate than Rosetta for \emph{small} ranges of $16\!\leq\!|R|\!\leq64$, which must probe larger area in its filters. The sudden fluctuations of Rosetta can be explained with the switch between different variants. The good FPR of \bloomRF{} for \emph{large ranges} (e.g., $10^{7}\!\leq\!|R|\!\leq\!10^{10}$) is due to the ability to probe more bits and the exact layer configurations. However, SuRF's LOUDS-encoding excels, for \emph{very large ranges} (e.g., $|R|\!=\!10^{11}$), while \bloomRF{} still achieves an acceptable FPR of 0.0454, as it probes larger areas of its mid-upper layers.
Under the same settings, we investigate the \emph{point-query} FPR (Fig.\ref{fig:exp1}.A2, B2 and C2 shown as figure-in-figure in Fig. \ref{fig:exp1}). 
Rosetta exhibits the lowest point-query FPR due to its accurate bottom filter-layer. \bloomRF{} needs more space for its mid-upper layers yielding slightly higher FPR. 
SuRF has the highest FPR due to its trie-truncation.
All PRF outperform Prefix-\BFs{} and fence pointers (Fig.\ref{fig:exp1}.D).
\textsf{Insight:} \bloomRF{} can handle a broad set of query ranges and outperforms all baselines, under various workload distributions, addressing Problem 1 (Sect. \ref{sec:intro}).

\begin{figure*}[!b]
	\begin{center}
		\includegraphics[width=\textwidth]{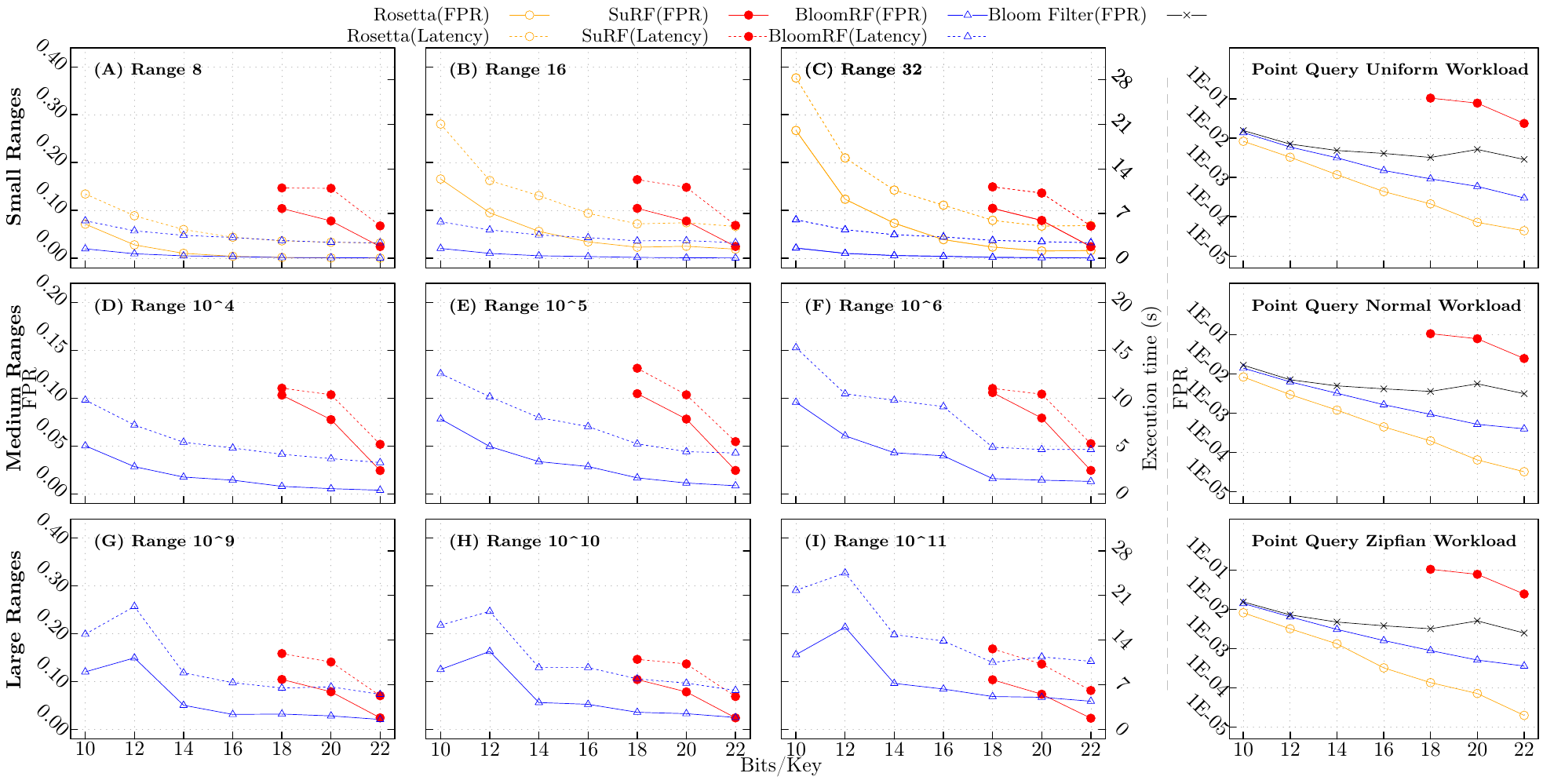}
		\caption{\bloomRFb{} is efficient, with better performance for different space budgets and query ranges in RocksDB.}
		\label{fig:exp2}
	\end{center}
\end{figure*}

\noindent{\bf Experiment 2: \bloomRFcaption{} is efficient.} 
We continue our comparison, by varying the space budget in RocksDB (Fig.\ref{fig:exp2}). We start from the 22 bits/key (favorable for all approaches and used in the previous experiment) and proceed to 10 bits/key, which is typical for standard \BFs{}. As we go, \emph{small} (Fig. \ref{fig:exp2}.A-C), \emph{medium} (Fig. \ref{fig:exp2}.D-F) and \emph{large} (Fig. \ref{fig:exp2}.G-I) range queries are performed. We use 50M keys; data and workload are uniformly distributed.

\bloomRF{} outperforms all baselines. It remains competitive to Rosetta for \emph{very small} ranges and bigger space budgets ($\geq\!18$ bits/key). \bloomRF{} also outperforms SuRF, except for \emph{very long} ranges ($|R|\!\geq\!10^{11}$). For \emph{point-lookups} in RocksDB (Fig. \ref{fig:exp2}, on the right)  \bloomRF{} is more accurate than the RocksDB \BF{} due to the random scatter and the error-correction. For point-queries and 2M keys, but in a \textit{standalone} setting (Fig. \ref{fig:exp4:7}.E1-E3) we compare all PRF, the Cuckoo-Filter \cite{Mitzenmacher:CuckooFilter:CONEXT:2014,cuckoo:lib} and the \BF{} from LevelDB~\cite{leveldb:lib}. We vary the  fingerprint sizes provided by the Cuckoo-Filter \cite{cuckoo:lib} and aim for high occupancies (95\%) to keep within the space budgets.

In terms of \textit{throughput} \bloomRF{} outperforms Rosetta 7\% to 44\% at 22 and 10 bits/key, respectively. We elaborate by providing a detailed breakdown of the probe-costs in RocksDB (Fig. \ref{fig:exp4:7}.G). We use 22 bits/key,  50M keys (2.06M per SST/filter), $10^{5}$ queries, uniform workload/data distribution. \bloomRF{} has the CPU- and total costs.
 
\textsf{Insight:} Considering the performance and FPR at smaller space budgets (Fig. \ref{fig:exp2}, $\leq\!18$ bits/key), we observe that \bloomRF{} is \emph{efficient}  in terms of: (i) performance per bits/key; and (ii) FPR per bits/key.

\begin{figure*}[!ht]
	\begin{center}
		\includegraphics[width=\textwidth]{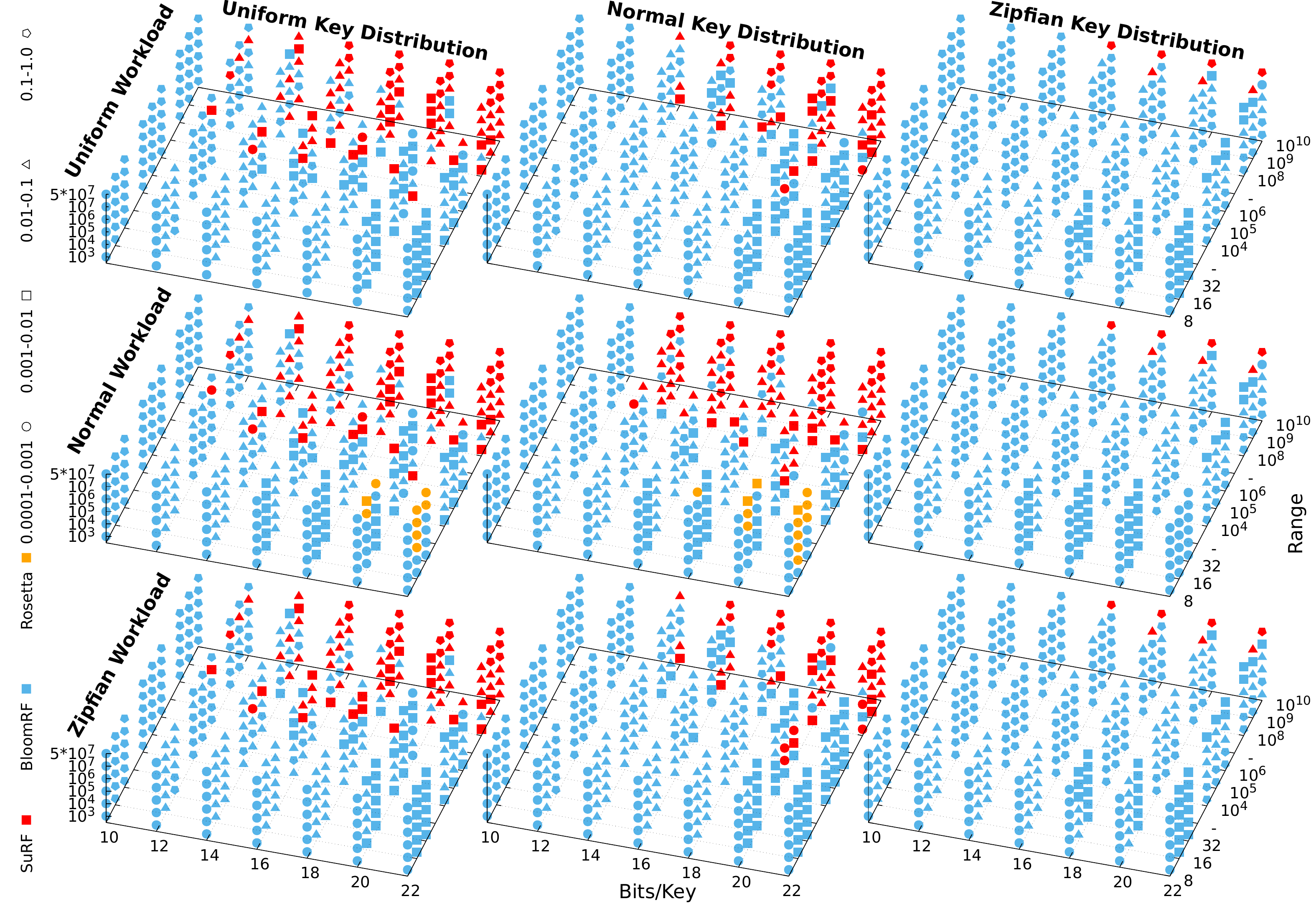}
		\caption{
		\bloomRFb{} handles different data and workload distributions and remains efficient for different space budgets (10..22 bits/key), query ranges (z-axis) and number of keys (y-axis). The color of each point represents the best filter, while the symbol stands for the relative FPR difference to the second best filter or to \bloomRFb{}, if not the best.}
		\label{fig:exp3}
	\end{center}
\end{figure*}

\noindent{\bf Experiment 3: \bloomRFcaption{} can handle skewed data distributions.} 
So far we only considered uniform data distributions. Now we relax this assumption and investigate the impact of \emph{normal} and \emph{zipfian} data distributions in a \emph{standalone} setting (Fig. \ref{fig:exp3}). We also vary the number of keys ($10^{3}$..50M), the space budget, the query range and the workload. The color of each point in Fig. \ref{fig:exp3} denotes the best filter, while the symbol stands for the FPR difference to the second best filter or to \bloomRF{}, in case it is not the best.

We observe that \bloomRF{} can handle \emph{skewed} data distributions across various settings. For zipfian \bloomRF{} is outperformed only in isolated cases. This is due to the underlying structure based on bloom-techniques, where bits from bottom-mid layers can be accurately probed due to its vertical error-correction, while SuRF is truncating beyond a certain length. Rosetta (presumably its hash functions or its variable-level design) loses efficiency with $|R|\!\geq\!16$.
 
\textsf{Insight:} Fig. \ref{fig:exp3} depicts a \emph{holistic} comparison among the PRF, on relevant parts of the problem space. All three approaches bring significant advantages to the design space and augment each other. 
Due to its LOUDS-encoding, SuRF tends to be better for large ranges ($10^{8}..10^{11}$), at higher space budgets with $\geq14$ bits/key and more keys. 
Rosetta tends to be better for very small query ranges with more than 16 bits/key. 
\bloomRF{} is generally applicable to various memory budgets, different number of keys, and performs well for different data distributions and workloads (Problem 3, Sec. \ref{sec:intro}).

\noindent{\bf Experiment 4: \bloomRFcaption{} is online and concurrent insertions have acceptable impact on its probe-performance at different insert/probe ratios.} 
We now quantify the online behavior, by investigating the impact of concurrent insertions on query performance and address Problem 2 (Sec. \ref{sec:intro}). To this end, we insert 50M, \emph{not sorted} or prepared, uniformly distributed keys with different (uniform) insert/lookup ratios (x-axis) in a \emph{standalone} setting. In single-threaded settings (Fig. \ref{fig:exp4:7}.A), the overall throughput increases with higher insert/lookup ratios. Hence, the impact of insertions is acceptable. 
A deeper analysis in multi-threaded settings (Fig. \ref{fig:exp4:7}.B) with varying the number of concurrent lookup/insertion-threads shows that insertions have marginal impact on the lookup performance per thread. The overall insert-throughput increases with more threads, although the throughput per insert-thread decreases. This is not surprising as \bloomRF{} is a parallel data structure. 

Next, we investigate the filter-construction costs (Fig. \ref{fig:exp4:7}.C) on the 50M, uniform dataset in RocksDB, where L0 comprises 25 SST files. We report the total creation and the serialization time (incl. tuning). \bloomRF{} has the lowest creation time, due to its high insertion performance. SuRF has relatively high overhead due to space budget tuning and trie creation.

\noindent{\bf Experiment 5: \bloomRFcaption{} can handle floats.} 
Our \emph{floating-point numbers} dataset  \cite{Kepler:2016}, contains positive and negative numbers. We execute 1.8M range queries (standalone), of size  $10^{-3}$. In absence of other baselines we only investigate \bloomRF{} (Fig. \ref{fig:exp4:7}.D).  In absence of other baselines we only show that \bloomRF{} achieves an avg. FPR of 0.18 for 10-22 bits/key and 4M lookups/s.

\noindent{\bf Experiment 6: \bloomRFcaption{} can serve as multi-attribute filter.} 
We evaluate multi-attribute querying in \bloomRF{} on a Sloan Digital Sky Survey DR16 \cite{sloan:SDSS:2019} dataset and extract the \emph{ObjectID} and the \emph{Run} columns. Their values roughly follow a normal distribution. In a standalone setting, we compare a multi-attribute \bloomRFparam{Run, ObjectID} probed with \textit{Run$<$300 AND ObjectID=Const} against two separate filters \bloomRFparam{Run} for \textit{Run$<$300} and \bloomRFparam{ObjectID} for \textit{ObjectID=Const}, combining the probe-results conjunctively. 

As shown in (Fig. \ref{fig:exp4:7}.F) \bloomRFparam{Run,ObjectID} yields better FPR than the combined FPR of the two separate filter-lookups \bloomRFparam{Run} and \bloomRFparam{ObjectID}. This observation is surprising since the separate filters operate on 64-bit integers, while the multi-attribute \bloomRF{} reduces precision and operates on 32-bit integers. The core intuition is that the FPR of \bloomRFparam{Run,ObjectID} depends on \textit{Z/Y}, where \textit{Y} and \textit{Z} are the number of data points satisfying \textit{ObjectID=Const} and \textit{Run$<$300 AND ObectID=Const} respectively.

\section{Related Work}
\label{sect:relwork}
Bloom-Filters are well-known and with many variants \cite{luo2019optimizing, Broder2005, Tarkoma:BloomFilters:2012, abdennebi2021bloom} covering different aspects: counting \cite{Mitzenmacher:CompressedBF:ESA:2006, Broder:TN:2000, Rottenstreich:CountingBF:INFOCOM:2012}; compressibility \cite{Mitzenmacher:CompressedBF:TN:2002}; SIMD vectorization \cite{Ross:SIMD:BF:DAMON:2014,Neumann:PErfOptimal:VLDB:2019}; partial deletes \cite{Rothenberg:DeletableBF:CL:2010}; efficient hashing \cite{DillingerM04, Mitzenmacher:Haashing:RSA:2008}; and data locality and novel hardware \cite{Canim:BufferPool:VLDB:2010, Debnath2011, Neumann:PErfOptimal:VLDB:2019, Guanlin:MSST:2011, Putze:JEA:2010}. Recently, there have been numerous novel proposals  \cite{Gupta:RAMBO:SIGMOD:2021,Dayan:Chucky:SIGMOD:2021,Pandey:VQF:SIGMOD:2021,Cole:ConditionalCockoo:SIGMOD:2021,Dayan:AdaptiveMerging:tods18}, all of which are point-filters with different properties. Pioneered by \cite{Kraska:LearnedIndex:SIGMOD:2018,Mitzenmacher:LearnedBF:NIPS:2018}, the concept of learned \BFs{}, leads to interesting applications \cite{vaidya2020partitioned,Liu:LearnedBF:VLDB:2020,Idreos:StackedFilters:2020} and is a future direction for \bloomRF{}.

The \emph{Adaptive Range Filter (ARF)} \cite{Alexiou:ARF:VLDB:2013} is one of the first approaches to describe the use of a simple form of dyadic numbering scheme to compute the covering intervals of a point. ARF, however, relies on a binary tree as a data structure and a powerful set of (learning) optimizations. Like \bloomRF{}, ARF relies on the concept of covering the whole domain of the datatype. \emph{SuRF} \cite{Zhang:SURF:SIGMOD:2018} shows the full potential of trie-based filters (Fast Succinct Trie) with a powerful encoding scheme (LOUDS-Dense/Sparse). In \bloomRF{} prefix hashing serves as an encoding scheme.

\emph{Rosetta} \cite{Dayan:Rosetta:SIGMOD:2020}, like \bloomRF{} utilizes \emph{\DIs{}} and \emph{dyadic decomposition}  for point-range-filtering. The concept itself is applicable to a wider range of other applications such as stream processing and summarization \cite{Cormode:CountMinSketch:JA:2005}, hot/cold data separation techniques \cite{Cormode:HotCold:TODS:2005} or persistent sketches \cite{Peng:persistentBF:SIGMOD:2018}.  
The \emph{Segment Trees} employed by  \cite{Dayan:Rosetta:SIGMOD:2020,Peng:persistentBF:SIGMOD:2018,Cormode:CountMinSketch:JA:2005} help encoding interval information and mapping range-queries into prefix-queries. \bloomRF{}'s prefix hashing achieves near space-optimal and computationally efficient encoding interval. 
Another major difference to \cite{Dayan:Rosetta:SIGMOD:2020,Peng:persistentBF:SIGMOD:2018,Cormode:CountMinSketch:JA:2005} is that \bloomRF{} employs PMHF to preserve local order. They reduce the number of memory accesses when  range querying and yields high range query performance.

\section{Conclusions}
\label{sect:conclusions}
We introduce \bloomRF{} as a unified PRF that extends  \BFs{} with range-lookups. We propose novel \textit{prefix hashing}  to encode range information in the hash-code of the key, and novel \textit{PMHF} for fast lookups and fewer memory accesses. We describe basic \bloomRF{} that is simple and tuning-free, and propose optimizations for handling larger ranges. \bloomRF{} has near-optimal space- and constant query-complexity and outperforms existing PRF by up to  4$\times$. 

\noindent{\bf Acknowledgments.} 
We thank the anonymous reviewers for the useful comments and suggestions. We are deeply grateful to the authors of \cite{Dayan:Rosetta:SIGMOD:2020} and \cite{Zhang:SURF:SIGMOD:2018} for providing the source code.

\balance

\clearpage
\newpage

\bibliographystyle{abbrv}
\bibliography{bloomRF-arXiv}

\balance

\end{document}